\documentclass[fleqn,usenatbib]{mnras} 

\usepackage{newtxtext,newtxmath}

\usepackage[T1]{fontenc}
\usepackage{lipsum}
\usepackage[normalem]{ulem}
\usepackage{xcolor}

\usepackage{graphicx}	
\usepackage{amsmath}	
\usepackage{multicol}
\usepackage{caption}
\usepackage{subcaption}
\usepackage{orcidlink}
\usepackage{xspace}

\DeclareRobustCommand{\VAN}[3]{#2}
\let\VANthebibliography\thebibliography
\def\thebibliography{\DeclareRobustCommand{\VAN}[3]{##3}\VANthebibliography}

\newcommand{\LSc}[1]{\textbf{\textcolor{orange}{[Lionel: #1]}}} 

\newcommand{\ryo}[1]{\textbf{\textcolor{cyan}{[ryo: #1]}}}

\newcommand{\nuc}{\mathrm{nuc}}
\newcommand{\dust}{\mathrm{dust}}
\newcommand{\gas}{\mathrm{g}}
\newcommand{\Msun}{M$_{\odot}$\xspace} 
\newcommand{\msun}{M$_{\odot}$\xspace}
\newcommand{\Rsun}{R$_{\odot}$\xspace} 
\newcommand{\Lsun}{L$_{\odot}$\xspace} 
 
\newcommand{\phant}{{\sc phantom}\xspace}

\newcommand{\mesa}{{\sc mesa}\xspace}

\title[Common envelope simulations with dust II.]{Dust formation in common envelope binary interactions -- II: \\ 3D simulations with self-consistent dust formation}

\author[Bermúdez-Bustamante et al.]{Luis C. Bermúdez-Bustamante\orcidlink{0000-0002-3629-6259 }$^{1,2}$ \thanks{E-mail: luiscarlos.bermudez@mq.edu.au}, 
    Orsola De Marco\orcidlink{0000-0002-1126-869X}$^{1,2}$,
Lionel Siess\orcidlink{0000-0001-6008-1103}$^3$, 
\newauthor
Daniel J. Price\orcidlink{0000-0002-4716-4235}$^4$, 
Miguel González-Bolívar \orcidlink{0000-0002-5939-9269 }$^{1,2}$, Mike Y. M. Lau \orcidlink{0000-0002-6592-2036}$^{5,6,4}$,
Chunliang Mu\orcidlink{0000-0003-1848-6507}$^{1,2}$,  \newauthor
Ryosuke Hirai\orcidlink{0000-0002-8032-8174}$^{4,6}$, Ta\"issa Danilovich\orcidlink{0000-0002-1283-6038}$^{4,7,8}$, and 
Mansi M. Kasliwal\orcidlink{0000-0002-5619-4938}$^9$
\\
$^1$School of Mathematical and Physical Sciences, Macquarie University, Balaclava Road, North Ryde, Sydney, NSW 2109, Australia\\
$^2$Astrophysics and Space Technologies Research Centre, Macquarie University, Balaclava Road, North Ryde, Sydney, NSW 2109, Australia\\
$^3$Institut d’Astronomie et d’Astrophysique, Université Libre de Bruxelles (ULB), CP 226, 1050 Brussels, Belgium\\
$^4$School of Physics and Astronomy, Monash University, Clayton, Victoria 3800, Australia\\
$^5$Heidelberger Institut f\"{u}r Theoretische Studien, Schloss-Wolfsbrunnenweg 35, 69118 Heidelberg, Germany \\
$^6$The ARC Centre of Excellence for Gravitational Wave Discovery (OzGrav2), Australia\\
$^7$The ARC Centre of Excellence for All Sky Astrophysics in 3 Dimensions (ASTRO 3D), Australia\\
$^8$Institute of Astronomy, KU Leuven, Celestijnenlaan 200D, 3001 Leuven, Belgium\\
$^9$Division of Physics, Mathematics, and Astronomy, California Institute of Technology, Pasadena, CA 91125, USA
}

\date{Accepted XXX. Received YYY; in original form ZZZ}

\pubyear{2023}

\defcitealias{GonzalezBolivar2023}{Paper~I}

\begin{document}
\label{firstpage}
\pagerange{\pageref{firstpage}--\pageref{lastpage}}
\maketitle

\begin{abstract}
We performed numerical simulations of the common envelope (CE) interaction between thermally-pulsing asymptotic giant branch (AGB) stars of 1.7~\Msun and 3.7~\Msun, respectively, and a 0.6~\Msun compact companion. We use tabulated equations of state to take into account recombination energy. For the first time, formation and growth of dust is calculated explicitly, using a carbon dust nucleation network with a C/O abundance ratio of 2.5 (by number). 
The first dust grains appear within $\sim$1--3~yrs after the onset of the CE, forming an optically thick shell at $\sim$10--20~au, growing in thickness and radius to values of $\sim$400--500~au over $\sim$40~yrs, with temperatures around 400~K.
Most dust is formed in unbound material, having little effect on mass ejection or orbital evolution.
By the end of the simulations, the total dust yield is $\sim8.4\times10^{-3}$~\Msun and $\sim2.2\times10^{-2}$~\Msun for the CE with a 1.7~\Msun and a 3.7~\Msun AGB star, respectively, corresponding to a nucleation efficiency close to 100\%, if no dust destruction mechanism is considered.
Despite comparable dust yields to single AGB stars, \textit{in CE ejections the dust forms a thousand times faster, over tens of years as opposed to tens of thousands of years}.
This rapid dust formation may account for the shift in the infrared of the spectral energy distribution of some optical transients known as luminous red novae. Simulated dusty CEs support the idea that extreme carbon stars and "water fountains" may be objects observed after a CE event.

\end{abstract}

\begin{keywords}
binaries: close -- stars: AGB and post-AGB -- stars: winds, outflows
--ISM: planetary nebulae
\end{keywords}

\section{Introduction}
\label{sec:introduction}

The CE binary interaction happens when an evolving, expanding star, such as a sub-giant, a red giant branch (RGB), an AGB, or a red supergiant star, transfers gas to a compact companion, either a lower mass main sequence star or a white dwarf, neutron star or black hole. Mass transfer tends to be unstable and leads to the formation of an extended common envelope surrounding the giant's core and the companion. As a consequence of the exchange of energy and angular momentum, the orbital separation is rapidly and greatly reduced. The envelope can be fully ejected, in which case the binary survives, or it can lead to a merger of the companion with the core of the giant. 

The CE interaction is the gateway to formation of compact evolved binaries like cataclysmic variables, X-ray binaries or the progenitors of supernova type Ia, and some black hole and neutron star binaries \citep{Paczynski1971,Ivanova2013,DeMarco2017}. Moreover, CE interactions are likely responsible for at least a fraction of intermediate luminosity transients \citep{Kasliwal2011}, such as luminous red novae \citep[e.g.,][]{Blagorodnova2017} and some of the latter are known to produce dust \citep{Tylenda2011,Nicholls2013}.

Advances in CE simulations have resulted in an initial understanding of the energetics that drive the expansion and eventual ejection of the envelope. 
While dynamical ejection of the entire envelope by transfer of orbital energy appears to be problematic \cite[but see][]{Valsan2023}, the work done by recombination energy liberated at depth shows promise as a way to enact full envelope unbinding \citep[e.g.,][]{Ivanova2015,Reichardt2020,Lau2022}.

Within this context, dust may provide an additional means to drive the CE, although it may be more or less effective depending on the type of star and interaction. Dust will also alter the opacities of the cooler parts of the CE ejecta, likely altering the appearance of the transient that results from the interaction. Past CE and other binary interaction simulations that considered dust were either 1D \citep{Lu2013}, analysed the conditions of the expanding envelope but did not calculate dust formation \citep{Glanz2018,Reichardt2019}, calculated dust formation in post-processing \citep{Iaconi2019a,Iaconi2020}, or worked with wider binary interactions \citep{Bermudez2020}. 

In a recent study \citep[][hereafter paper I]{GonzalezBolivar2023} we carried out two 3D CE simulations, with a 1.7~\msun\ and a 3.7~\msun\ AGB donor star respectively, where the dust opacity was estimated using the simplified prescription devised by \citet{Bowen1988}. While the Bowen approximation has been tested for the case of single AGB stars \citep[e.g.][]{Chen2020,Bermudez2020,Esseldeurs2023}, it has not been used before in CE calculations. In \citetalias{GonzalezBolivar2023} we found that dust driving has a limited effect on the unbound envelope mass. However, the opacity of the cool, expanded envelope is greatly increased by the dust. 

There are several novelties in the present work, compared to \citetalias{GonzalezBolivar2023}: first and foremost, instead of focusing on traditional CE diagnostics like envelope ejection dynamics, this study centers on the intricate process of dust nucleation. The diagnostics serve solely to provide context and facilitate comparison between the dusty results and the non-dusty simulations.
Second, we calculate dust opacity directly, whereas in \citetalias{GonzalezBolivar2023}, the opacity was estimated based on the simplified Bowen approximation, which does not provide detailed information on dust grain properties such as mass, size, or opacity. This study addresses that gap by offering comprehensive data on these properties.
Third, we also investigate the effects of dust on observational signatures, a topic we could only speculate about in our previous work. This is crucial for accurately calculating light curves from 3D simulations, as the dust component plays a vital role. While the Bowen method allows for initial estimates, it is not suitable or reliable for generating realistic light curves.

The paper is organised as follows: in Section~\ref{sec:setup} we describe the initial conditions of the binary system, together with the dust formation process and the calculation of the dust opacity. The dust properties and the evolution of the binary system, are presented in Section~\ref{sec:resultados} alongside the size and shape of the photosphere. The results are discussed in Section~\ref{sec:Discussion} and conclusions are presented in Section~\ref{sec:Conclusions}.
\section{Simulation Setup}
\label{sec:setup}
\begin{figure}
    \centering
    \includegraphics[width=1.0\linewidth]{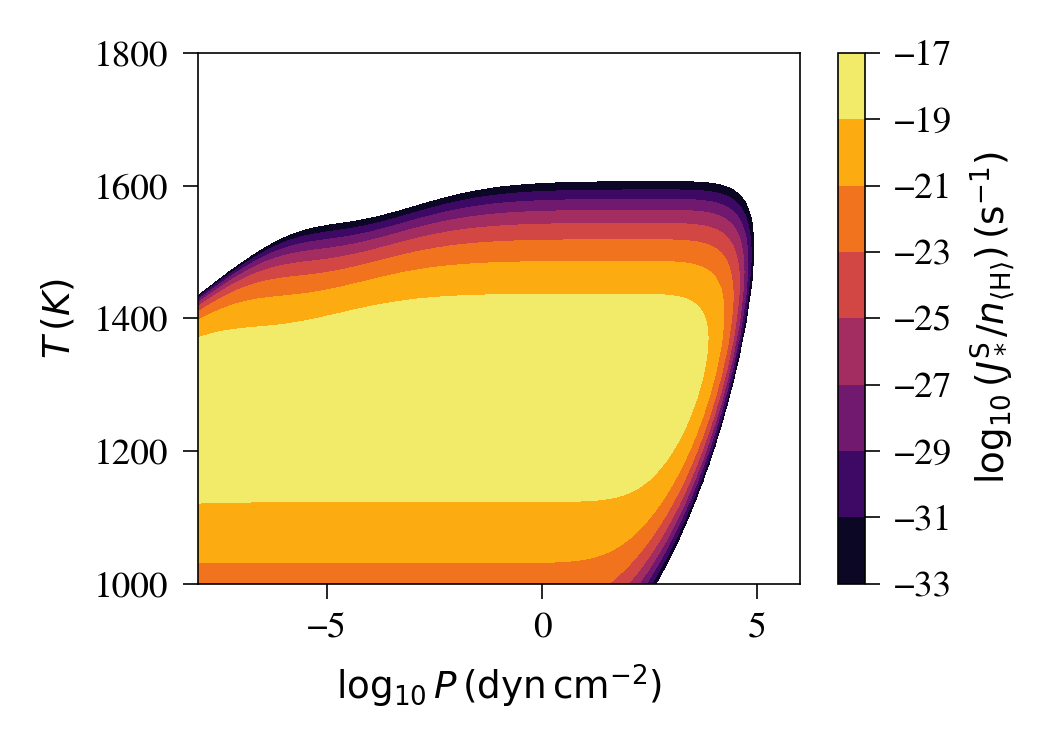}
    \caption{The nucleation rate per hydrogen atom $(\widehat{J}^S_*=J^S_*/n_{\langle H \rangle})$ as a function of temperature and pressure, for a C/O number ratio of 2.5.}
    \label{fig:nucleation_rate}
\end{figure}

\subsection{Initial Conditions}
For our dusty CE simulations, we use two stellar models that were calculated using the one-dimensional implicit stellar evolution code Modules for Experiments in Stellar Astrophysics (MESA), version 12778 \citep{Paxton2011,Paxton2013,Paxton2015}.
The first model is the same one used by \cite{GonzalezBolivar2022}, namely an AGB star with a mass of 1.71~\msun{} (zero-age main sequence mass of 2~\msun), a radius of 260~\Rsun, a CO-rich core of 0.56~\Msun, a metallicity $\rm Z = 0.021$, and at its seventh thermal pulse.
The second stellar model comes from a zero-age main sequence star with a mass of 4~\Msun and metallicity $\rm Z = 0.024$ that was evolved to the third AGB thermal pulse, at which point it has a mass of 3.7~\msun, a radius of 343~\Rsun{}, and a core mass of 0.72~\Msun (for more details, see \citetalias{GonzalezBolivar2023}).
In mapping the MESA models to \phant, the core of each AGB star was replaced by a sink particle of similar mass and with a softening radius $r_\mathrm{soft}$ of 2.5 and 8.0~\Rsun for the 1.7 and 3.7~\Msun model, respectively.
Both models are placed in orbit with a 0.6~\Msun companion star, which is modelled as a point particle and may represent either a main-sequence star or a white dwarf.
The simulation starts at the onset of Roche lobe overflow (inputs and outputs of the simulations are listed in Table~\ref{tab:simulations}).

\begin{table*}
    \centering
    \begin{tabular}{lcccccccc}
     \hline
     Model & $M_1$ & $q$ & Dust opacity & $r_\mathrm{soft}$ & $a_i$& $a_\mathrm{f}$ & $M_{\rm ub}$ & $M_{\rm dust}$\\
           & (\Msun) &&\bf{($cm^2~g^{-1}$)}& (\Rsun) & (\Rsun/au)& (\Rsun/au) & (\Msun/\%) & (\Msun)\\
     \hline
     Non-dusty 1.7  & 1.7 &0.35& No & 2.5 &550 / 2.6&33 / 0.15 &0.93 / 82 & --\\
     Bowen 5 1.7    & 1.7 &0.35& Bowen $\kappa_{\rm max} = 5$ & 2.5&550 / 2.6&33 / 0.15&0.96 / 84 & --\\
     Bowen 15 1.7   & 1.7 &0.35& Bowen $\kappa_{\rm max} = 15$ & 2.5&550 / 2.6&40 / 0.19 &1.01 / 88 & --\\
     Dusty 1.7      & 1.7 &0.35& Nucleation & 2.5 &550 / 2.6&42 / 0.20 &1.00 / 88 & $8.2\times10^{-3}$\\
     Non-dusty 3.7  & 3.7 &0.16& No & 8.0 &637 / 3.0&10 / 0.05&2.47 / 83 & --\\
     Bowen 5 3.7    & 3.7 &0.16& Bowen $\kappa_{\rm max} = 5$&8.0&637 / 3.0&10 / 0.05&2.35 / 79 & --\\
     Bowen 15 3.7   & 3.7 &0.16& Bowen $\kappa_{\rm max} = 15$&8.0&637 / 3.0&11 / 0.05&2.56 /  86 & --\\
     Dusty 3.7      & 3.7 &0.16& Nucleation & 8.0 &637 / 3.0&12 / 0.06&2.48 / 83 & $2.2\times10^{-2}$\\
     \hline
    \end{tabular}
    \caption{Characteristics of the simulations. In all cases, the companion has a mass of $M_2 = 0.6$~\Msun. The 1.7~\Msun\ (3.7~\Msun) star has a radius of 260~\Rsun (343~\Rsun), and a core mass of 0.56~\Msun (0.72~\Msun). From left to right, the columns indicate (1) the name of each simulation, (2) the mass of the giant donor star (3) the mass ratio, $q=M_2/M_1$ (4) whether dust is accounted for and the method used to calculate its opacity, (5) the softening length of the (core) point mass particles, (6) the initial and final (7) separation (in \Rsun\ and au) at 12.5 years. The eighth column is the unbound mass by the mechanical criterion (see text) at 12.5 years (in \Msun\ and as percentage of the envelope mass -- 1.14~\Msun\ and 3.01~\Msun\ for the 1.7~\Msun\ and 3.7~\Msun\ models, respectively). The last column gives the dust mass at the end of the simulations ($\sim 44$~years).}
    \label{tab:simulations}
\end{table*}

For all simulations, we use the smoothed particle hydrodynamics (SPH) code \textsc{phantom} \citep{Price2018} with $1.37\times10^6$ SPH particles and \phant 's implementation of \mesa 's OPAL/SCVH equation of state tables \citep[as done by][]{Reichardt2020} to account for recombination energy in this adiabatic simulation.
Approximately 8.6 years after the start of the 1.7~\Msun model, we have modified the simulation setup, specifically we have changed the sink particle accretion radius from 0~\Rsun (i.e., no accretion) to 15~\Rsun. The reason behind the change at that particular instant is to "remove" $\sim$6000 particles (less than 0.5\% of the total SPH particles) positioned within that distance from the companion star. The density of those SPH particles is very high, and their smoothing length is very small \citep[Eq. 6 in][]{Price2018}, which causes the time step in the simulation to decrease to such a magnitude that the computational cost is prohibitively high \citep[see section 2.3.2 in][]{Price2018}. By removing those particles, the time step increases and allows the simulation to be concluded at a reasonable cost. This change in configuration does not alter the validity of our results, as we are interested in the formation of dust in the ejected material, which occurs far away from the center of mass, as shown in the next section. 
A study of both resolution and conservation of energy and angular momentum is provided in Appendix~\ref{AppendixB} to support our claim.  
It was not necessary to remove SPH particles in the 3.7~\Msun model.

\subsection{Dust formation}
We consider the formation of carbon-rich dust assuming a C/O abundance ratio of 2.5 (by number) in the gas phase. Even though the AGB models had a C/O ratio equal to 0.32, the structure of the convective envelope is not expected to be significantly affected by this change in C content. 
The choice of C/O=2.5 corresponds to the maximum C/O ratio obtained by \cite{Ventura2018} for AGB stars in the main sequence mass range 1.5-3.5~\Msun\ (see their Fig. 14). Using this high value will provide an upper limit to the amount of carbon dust that can be produced.

The dust making process takes place in two steps: i) the nucleation stage, in which seed particles from the gas phase are formed, followed by ii) the growth phase, where monomers (dust building blocks) accumulate on the seed particles to reach macroscopic dimensions. As a result, dust formation depends on the abundance of carbon-bearing molecules in the gas, which are obtained by solving a reduced chemical network for the carbon-rich mixture. 
Here we give a brief account of the salient points of the nucleation formalism so that the reader can understand the following analysis (for more details, refer to \cite{Siess2022}, \cite{Gail2013}, and \citet[][and subsequent papers]{Gail1984}). This formalism is based on the evolution of the moments, $\mathcal{K}_i$, of the grain size distribution $f(N,t)$. For  order $i$, the moment can be written as:  
\begin{equation}
    \mathcal{K}_i = \sum_{N=N_l}^\infty N^{i/3} f(N,t),
\end{equation}
where $N$ is the number of monomers in a grain (in our case C, C$_2$, C$_2$H and C$_2$H$_2$) and $N_l \sim 1000$ is the minimum number of monomers that a cluster must contain to be considered as a dust grain. 
The knowledge of the moments gives direct access to the dust properties, for example the average grain radius: 
\begin{equation}
    \langle r\rangle  = a_0\frac{\mathcal{K}_1}{\mathcal{K}_0},
\end{equation}
where $a_0 = 1.28\times 10^{-4}$~$\mu$m is the radius of a monomer (namely that of a carbon atom) inside a dust grain.
If we considered the normalized variables $\widehat{{\cal K}_i} = {\cal K}_i/n_{\langle \mathrm{H}\rangle}$ and $\widehat{J}_* = J_*/n_{\langle \mathrm{H}\rangle}$, where $J_*$ is the nucleation rate of seed particles per unit volume, and $n_{\langle \mathrm{H}\rangle}$ is the number of hydrogen atoms per unit volume, the equations governing the evolution of the moments can be written as:
\begin{eqnarray}
    \label{jevol}
    \frac{\mathrm{d}\widehat{J}_*}{\mathrm{d}t} & = &\frac{\widehat{J}_*^s - \widehat{J}_*}{\tau_*}, \\
    \label{moment0}
    \frac{\mathrm{d}\widehat{\mathcal{K}}_0}{\mathrm{d}t} & = &\widehat{J}_*, \\
    \label{momenti}
    \frac{\mathrm{d}\widehat{\mathcal{K}}_i}{\mathrm{d}t} & = & \frac{i\,  \widehat{\mathcal{K}}_{i-1}}{3 \tau} +N_l^{i/3}\widehat{J}_*,
\end{eqnarray}
where $\widehat{J}_*^s$ is the quasi-stationary nucleation rate per hydrogen atom, $\tau_*$ is the relaxation time of nucleation towards equilibrium and $\tau$ is the growth timescale  ($>0$) or evaporation timescale ($<0$) of the grains \citep[for details, see][]{Siess2022}. It should be emphasized that in our current implementation, dust is not destroyed and can only grow.  
The dependence of the  nucleation rate  on temperature and pressure for our adopted C/O = 2.5 is illustrated in  Figure~\ref{fig:nucleation_rate}: it peaks around 1100-1400~K over a wide range of pressures, $P$, from $\sim10^{-8}$ to $\rm \sim10^{3}~dyn\,cm^{-2}$.

Another important parameter is the supersaturation ratio, $S$, given by
\begin{equation}
S = \frac{P_\mathrm{C}(T_\mathrm{g})}{P_\mathrm{sat}}, 
\label{eq:def_S}
\end{equation}
where $P_\mathrm{C}(T_\mathrm{g})$ is the partial pressure of carbon in the gas phase, $P_\mathrm{sat}$ is the vapor saturation pressure of carbon in the solid phase and $T_\mathrm{g}$ is the gas temperature. For nucleation to occur, $S$ must be larger than a critical value $S_c$, of the order of unity, and this happens when the temperature drops below $\sim 1500$~K\footnote{Note that in regions of very low density where dust formation is inefficient, $S$ can be much larger than unity because the value of $P_{\rm sat}$ is very small.}.

The dust opacity, $\kappa_\mathrm{d}$, is given by
\begin{equation}
    \kappa_\mathrm{d} = \frac{\pi a_0^3}{\rho}Q^\prime_\mathrm{ext}{\mathcal{K}_3},
    \label{eq:kappa_d}
\end{equation}
where the expression $Q^\prime_\mathrm{ext} = 6.7\, (T_\mathrm{d}/\mathrm{K})$~cm$^{-1}$ is a fit to the Planck mean of the Mie extinction coefficient (Eq. 13 of \citealt{Draine1981}), and $\rho$ is the gas density.
The value of $Q^\prime_\mathrm{ext}$ depends only on the dust temperature, $T_\mathrm{d}$,  and is strictly valid for grains of size $a \lesssim 0.1\mu m$ (for details, see Sect 3.2.2 of \citealt{Siess2022} and Sect 7.5.5 of \citealt{Gail2013}). In the absence of a proper radiative transfer treatment in our simulations, we assume that gas and dust are thermally coupled, i.e., $T_\mathrm{g} = T_\mathrm{d} = T$. The gas opacity is set to a constant value of $\kappa_\gas = 2\times$10$^{-4}$~cm$^2$~g$^{-1}$, which is a good approximation at temperatures below $10^4$\,K, where hydrogen is mostly recombined \citep{Siess2022}.

The dust mass is the sum, over all the SPH particles, of the average mass of carbon atoms condensed into dust, and is given by

\begin{equation}
\mathrm{M}_{\rm dust} = \sum_\mathrm{part.} \frac{m_\mathrm{SPH}}{\bar{m}_{\mathrm{H}}} m_c\,\widehat{\mathcal{K}}_3,
\label{eq:dust_mass}
\end{equation}
where $\widehat{\mathcal{K}}_3$ gives the average number of condensed carbon atoms per H-atoms for each SPH particle, $m_c$ is the mass of a carbon atom and $m_{\mathrm{SPH}}$ is the (constant) mass of an SPH particle. The mean mass per hydrogen atom, $\bar{m}_{\mathrm{H}}$, is given by 
\begin{equation}
    \bar{m}_{\mathrm{H}} = m_u \,\sum_i\,A_i\,\epsilon_i,
\end{equation}
where $\epsilon_i$ is the abundance by number relative to hydrogen, $A_i$ is the atomic weight of chemical element $i$ and $m_u$ the atomic mass unit.
Similarly, the maximum theoretical dust mass, when all carbon atoms that are not locked in CO molecules are in the grains, can be estimated as
\begin{equation}
\mathrm{M}_{\rm carbon} = \frac{m_c}{\bar{m}_{\mathrm{H}}}\,(\epsilon_\mathrm{C}\,-\epsilon_\mathrm{O})\sum_\mathrm{part.} m_\mathrm{SPH} = \frac{m_c}{\bar{m}_{\mathrm{H}}}\,(\epsilon_\mathrm{C}\,-\epsilon_\mathrm{O}) \mathrm{M_\mathrm{env}}, 
\label{eq:maximum_carbon}
\end{equation}
where $\epsilon_\mathrm{C}$ and $\epsilon_\mathrm{O}$ are the abundance of carbon and oxygen by number relative to hydrogen and $\rm M_\mathrm{env}$ is the envelope mass.

Each SPH particle has an initial composition set by $\epsilon_i$. As nucleation proceeds, carbon atoms are progressively agglomerated into the dust, but the knowledge of $\widehat{\mathcal K}_3$ allows us to determine, at each time step, the number of carbon atoms remaining in the gas phase. So the chemical composition and dust content of each SPH particle vary during the calculations, but the particle mass remains constant because the SPH particle always contains the same number of atoms, whether they are free, or locked in molecules or dust.

The knowledge of the dust opacity allows us to calculate the radiative acceleration\footnote{Inside the star, the outward radiative force is several orders of magnitude smaller than the inward gravity force, therefore the hydrostatic equilibrium of the star is not disturbed.}  
on the particles as an additive term to the equation of motion, which becomes
\begin{equation}
    \frac{d\mathbf{v}}{dt} = -\frac{ \nabla P}{\rho} - \nabla\phi_\mathrm{gas-gas} - \nabla\phi_\mathrm{sink-gas} + \frac{(\kappa_\gas+ \kappa_\mathrm{d})L}{4\pi r^2 c}\frac{\mathbf{r}}{r},
    \label{eq:motion}
\end{equation}
where $P$ is the pressure, $\phi_\mathrm{gas-gas}$ and $\phi_\mathrm{sink-gas}$ are the gravitational potentials from the SPH particles (self-gravity) and between a particle and the sink masses, respectively. The variable $r$ is the distance from the AGB core, which is assumed to produce a constant luminosity $L = 5180$~L$_\odot$ for the 1.7~\Msun model and $L = 9010$~L$_\odot$ for the 3.7~\Msun model. For regions where dust cannot form, the last term in the equation of motion has negligible impact, since we assume a gas opacity $\kappa_\gas = 2\times 10^{-4}$~$\rm cm^2 g^{-1}$.

\section{Results}
\label{sec:resultados}

\begin{figure*}
    \includegraphics[width=0.49\linewidth]{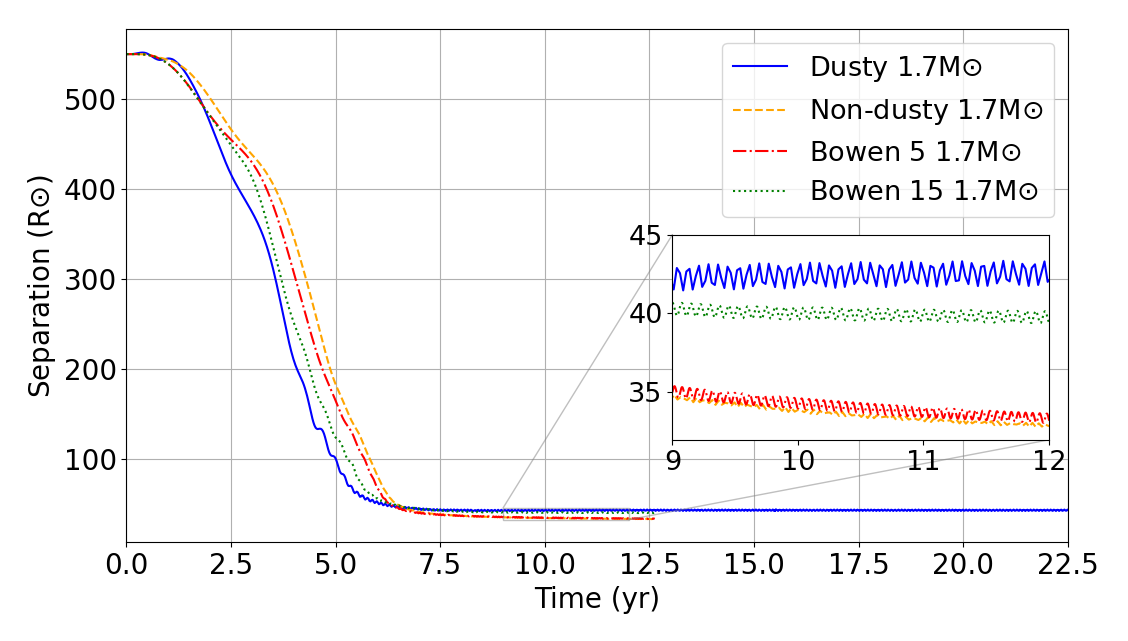}
    \includegraphics[width=0.49\linewidth]{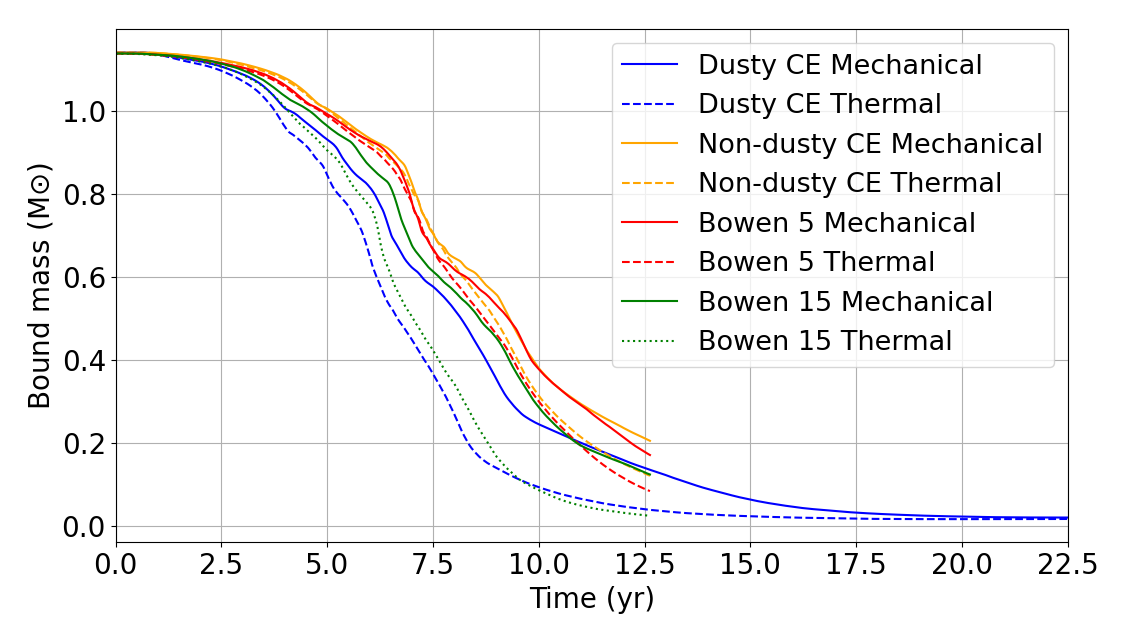}\\
    \includegraphics[width=0.49\linewidth]{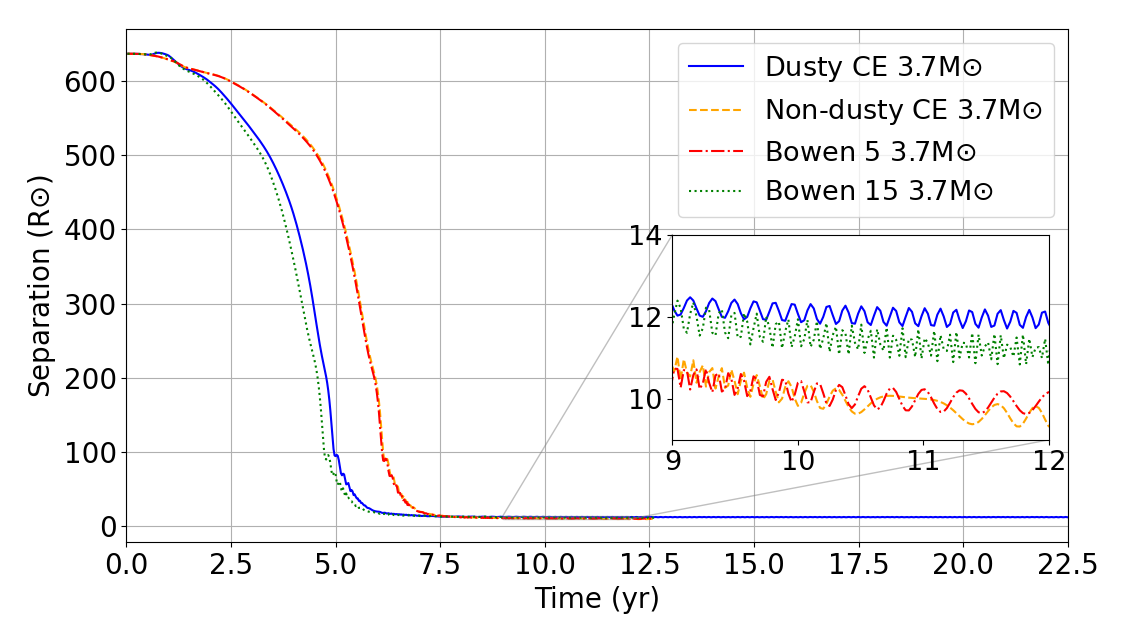}
    \includegraphics[width=0.49\linewidth]{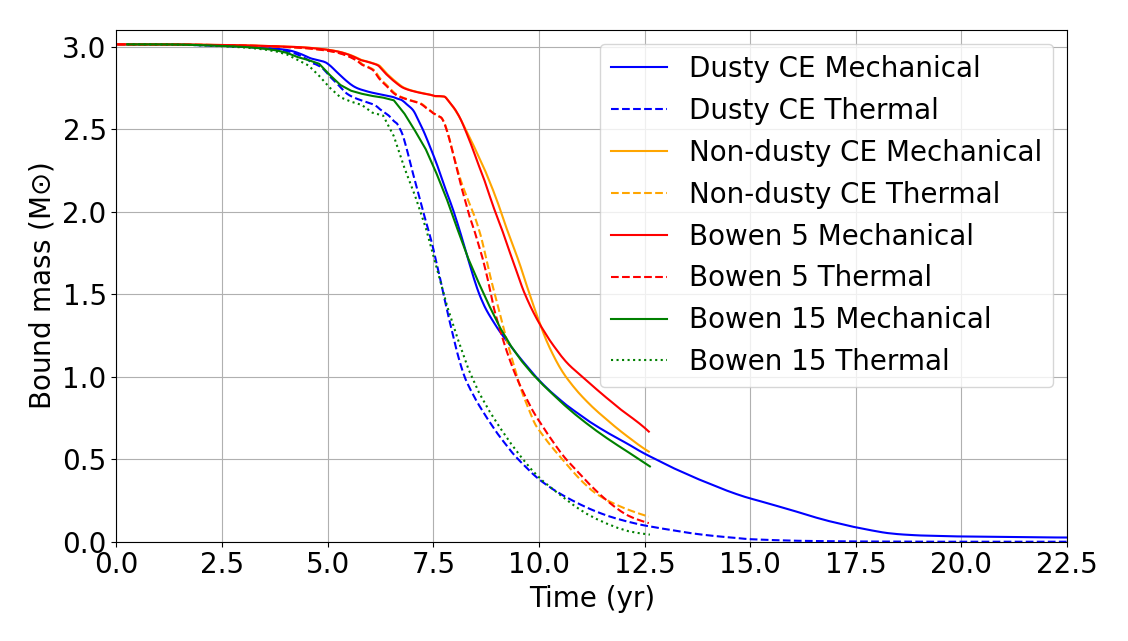}
    \caption{Orbital separation (left panels) and bound mass (right panels) as a function of time for the CE simulations with a 1.7~\msun\ (top) and a 3.7~\msun\ (bottom) AGB primary star. Simulations with dust nucleation are shown in blue, those using the Bowen's formulation in red (green) when the maximum opacity $\kappa_\mathrm{max}=5~\mathrm{cm^2\,g^{-1}}$ ($\kappa_\mathrm{max}=15~\mathrm{cm^2\,g^{-1}}$). CE simulations without dust are shown in yellow. The insert in the left panels shows the  orbital separation at the end of the in-spiral phase. The mechanical and thermal bound masses are displayed in the right panels with the solid and dashed lines, respectively.}
    \label{fig:separation_massunbound}
\end{figure*}

In this section we describe where and when dust forms in the envelope, analyse the spatial distribution of the mean grain size and the evolution of the dust mass. To anchor the dust formation processes to the in-spiral timeline, we anticipate the discussion of Sect.~\ref{ssec:orbital-evolution} and present in Fig.~\ref{fig:separation_massunbound},  the evolution of the separation and unbound mass. After comparisons with previous simulations,  we investigate in Sect~\ref{sec:photo} the expected size of the CE photosphere.

\subsection{Dust Properties}
\label{ssec:dust-properties}

\subsubsection{Nucleation, growth and opacity of dust grains}

\begin{figure*}
    \centering
    \includegraphics[width=0.3\linewidth]{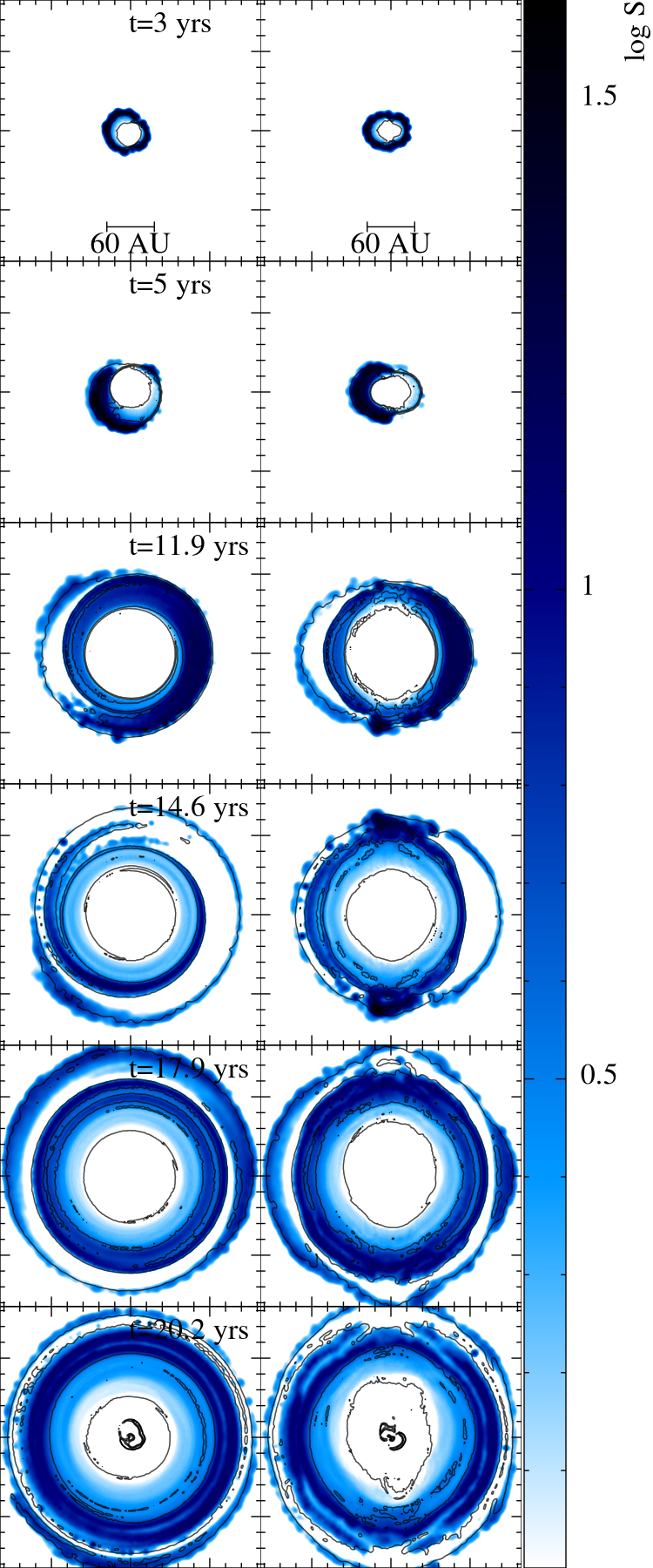}
    \hspace{0.7cm} 
    \includegraphics[width=0.3\linewidth]{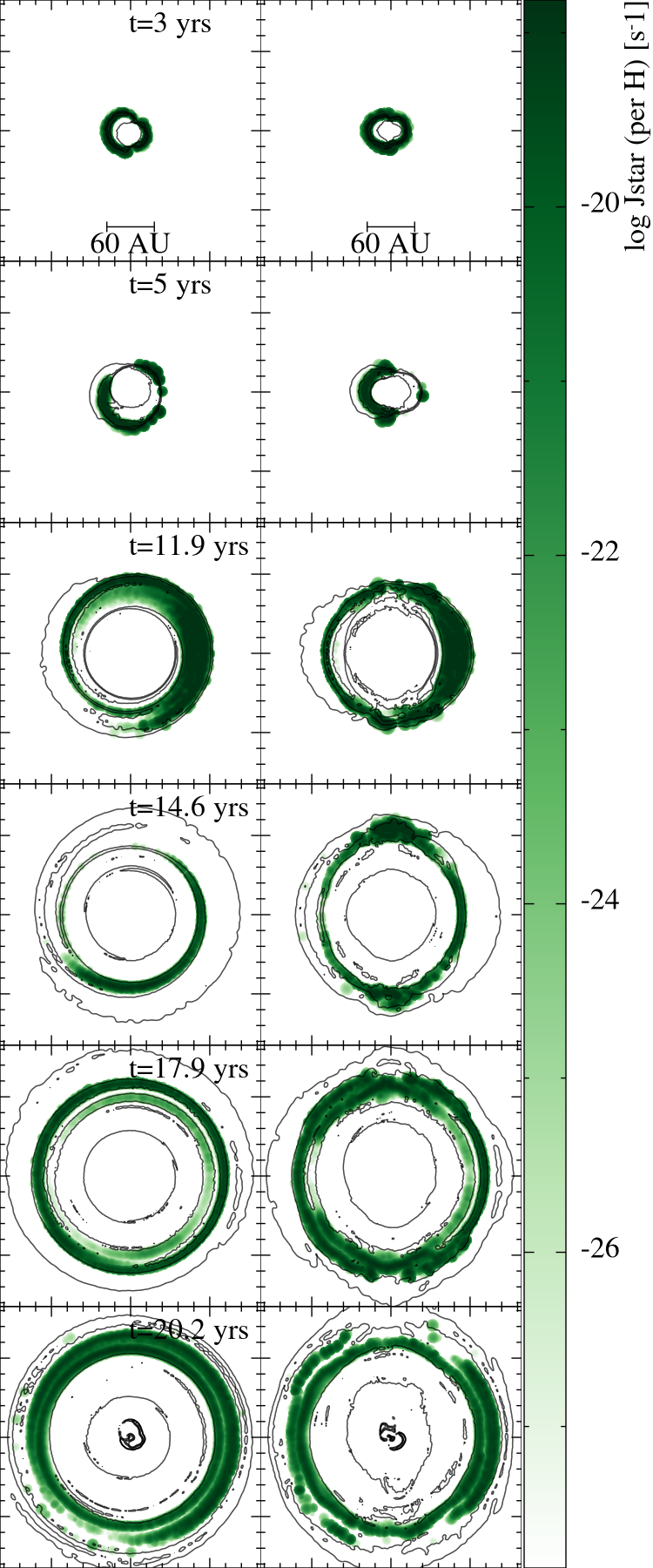}
    \hspace{0.7cm} 
    \includegraphics[width=0.3\linewidth]{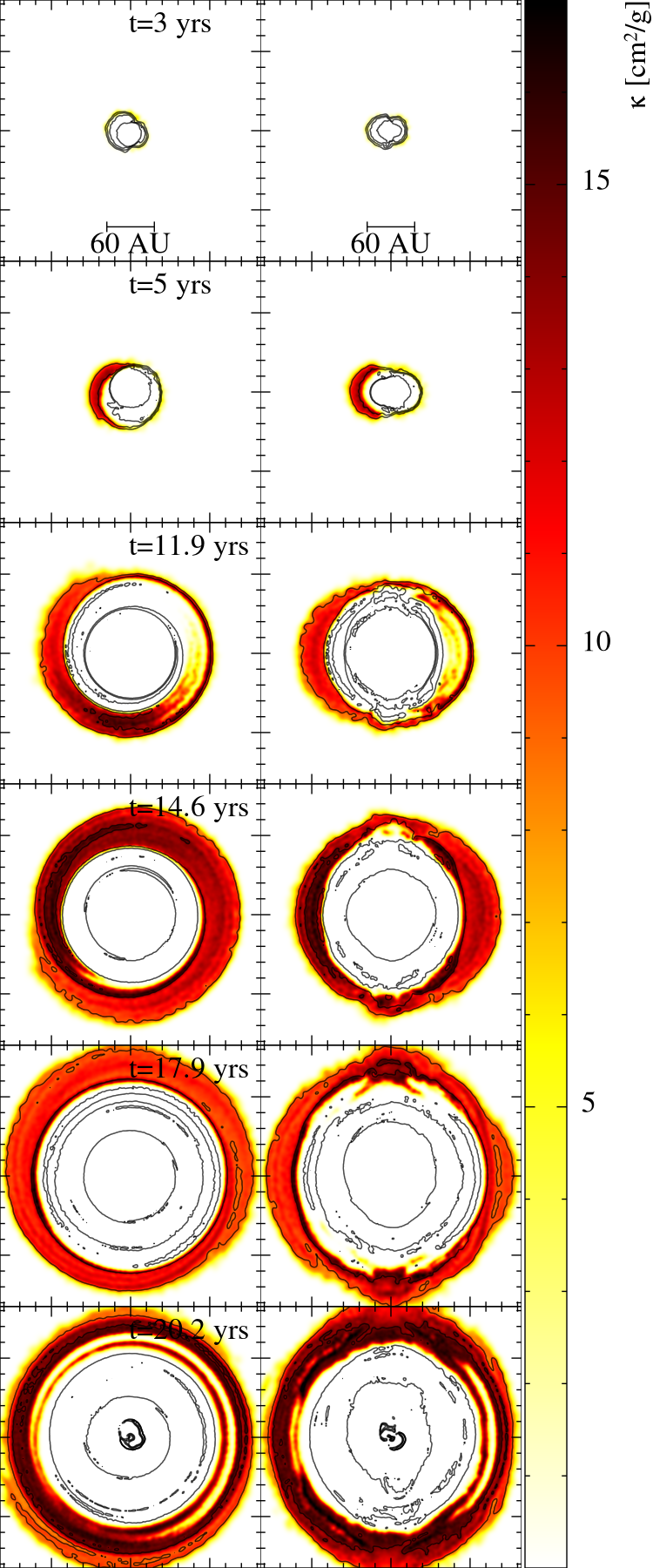} 
    \caption{Slices in the XY (columns 1, 3 and 5) and XZ (columns 2, 4 and 6) planes of supersaturation ratio, $S$ (blue), nucleation rate per hydrogen atom, $\hat{J}_{\star}$ (green) and opacity, $\kappa$ (red) at 6 different times for the 1.7~\msun\ model. Black temperature contours are equally spaced in linear scale and range between 750~K (outermost contour) to 2000~K (innermost contour). Each panel is 160 by 160 au per side. All figure movies can be found at the following URL: \url{https://tinyurl.com/y455avdj}} 
    \label{fig:S_J_kappa_2Mo}
\end{figure*}

\begin{figure*}
    \centering
    \includegraphics[width=0.3\linewidth]{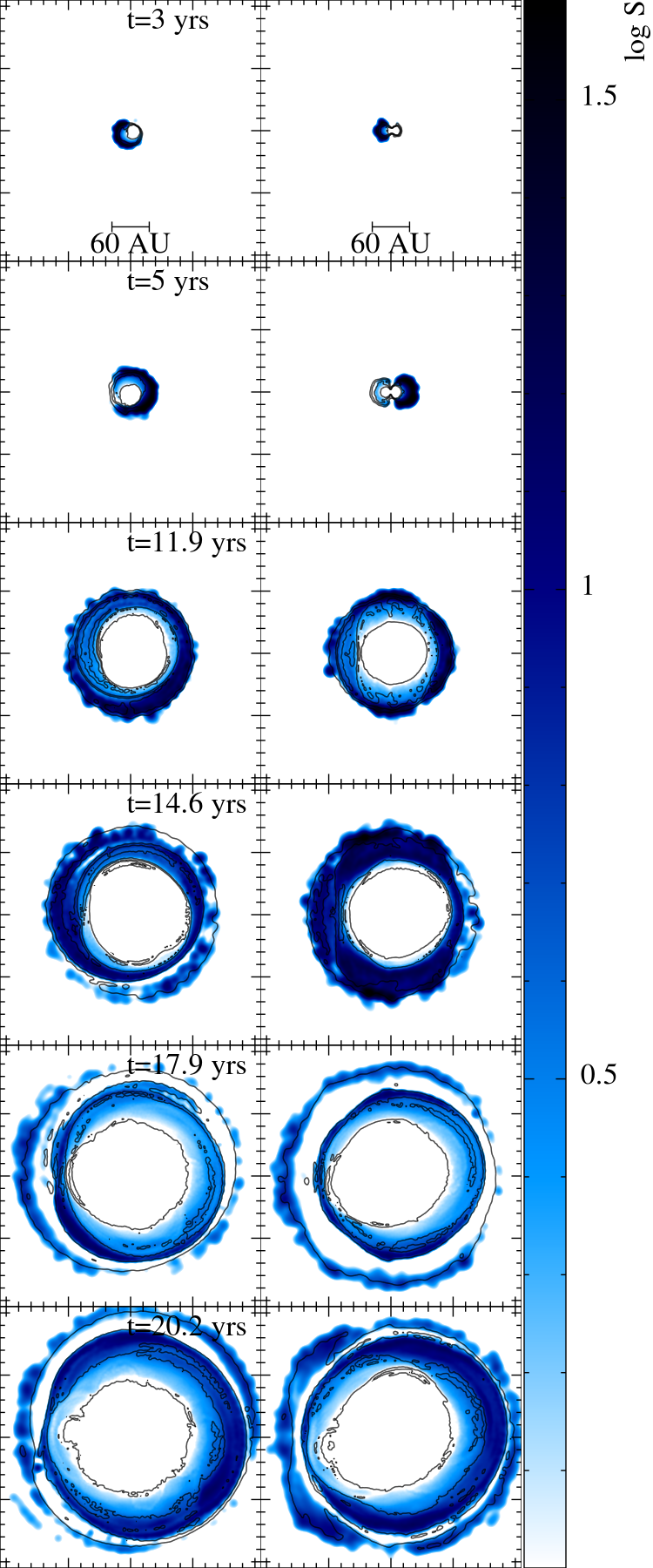}
    \hspace{0.7cm}
    \includegraphics[width=0.3\linewidth]{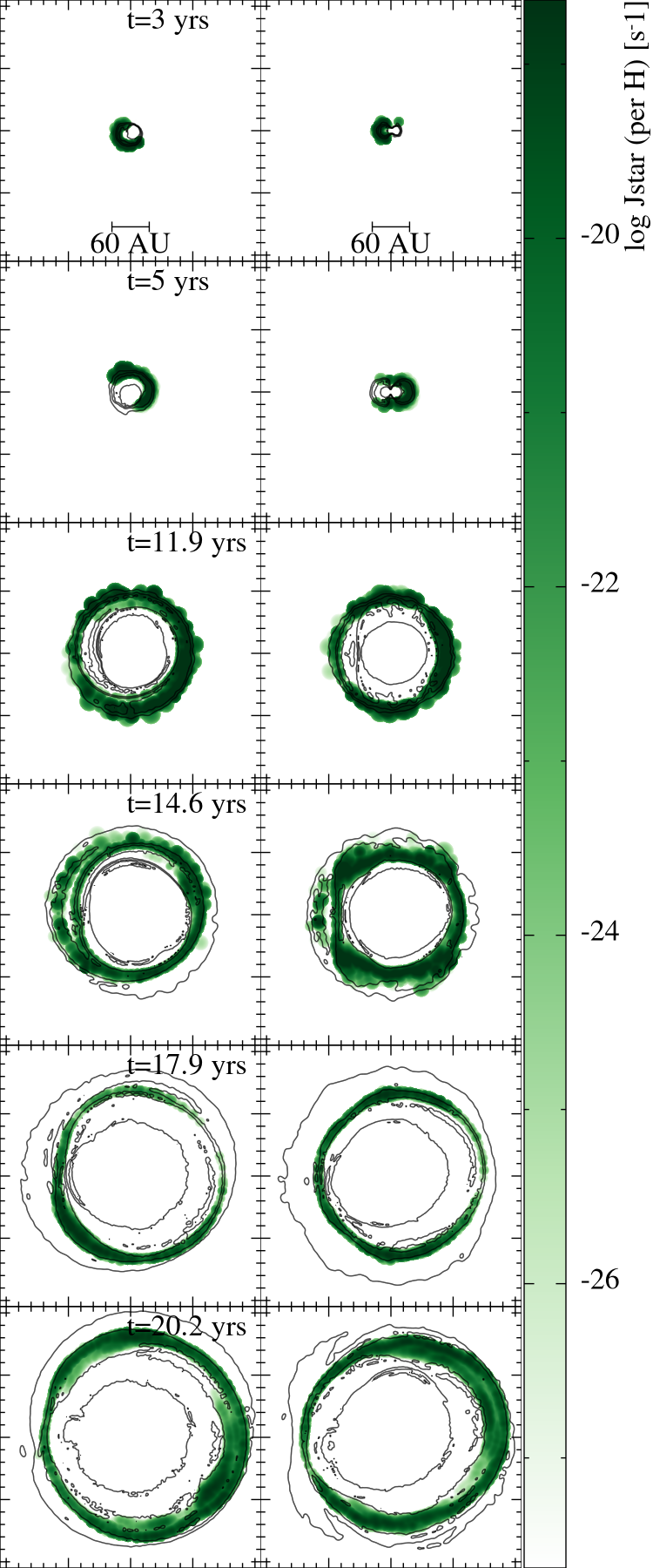}
    \hspace{0.7cm}
    \includegraphics[width=0.3\linewidth]{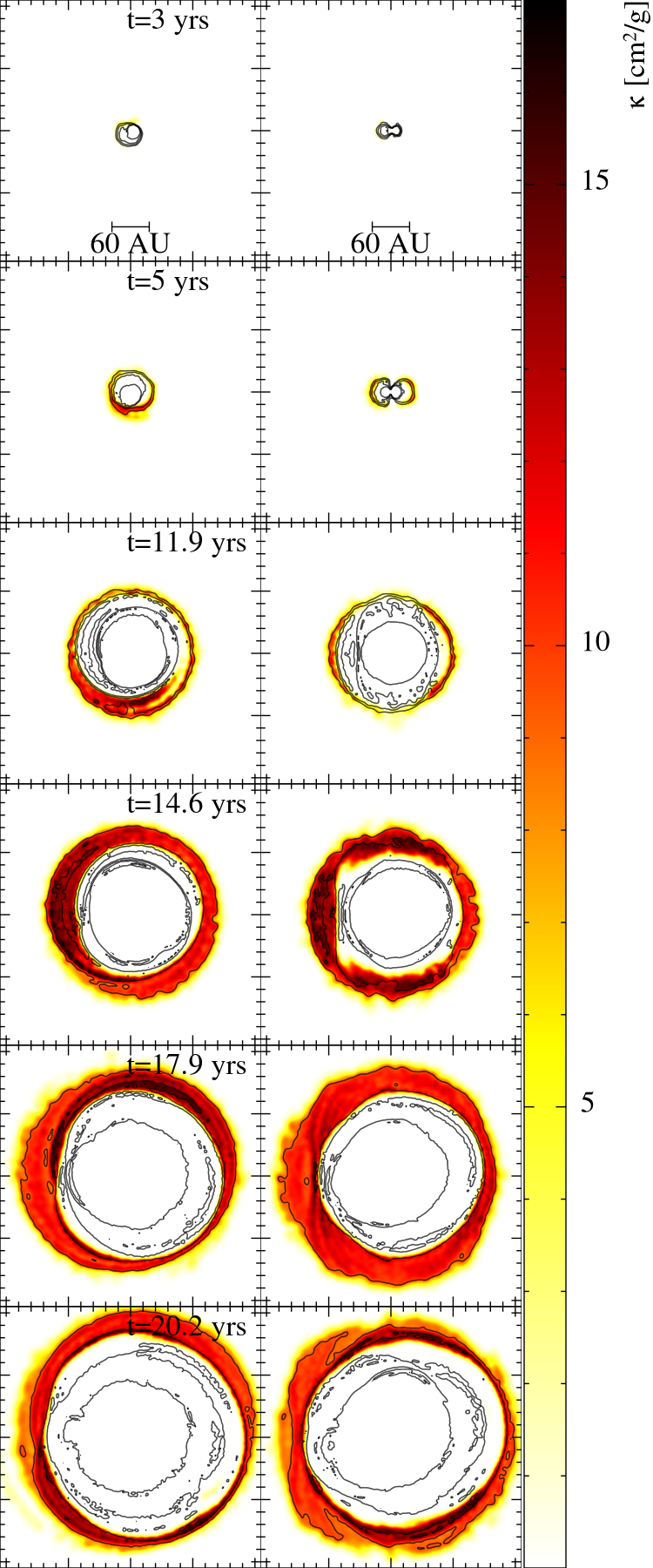}      
    \caption{ Slices in the XY (columns 1, 3 and 5) and XZ (columns 2, 4 and 6) planes of supersaturation ratio, $S$ (columns 1 and 2), nucleation rate per hydrogen atom, $\hat{J}_{\star}$ (columns 3 and 4) and opacity, $\kappa$ (columns 5 and 6) at 6 different times for the 3.7~\msun\ model. The temperature contours and the scale bar are the same as in Figure~\ref{fig:S_J_kappa_2Mo}. All figure movies can be found at the following URL: \url{https://tinyurl.com/y455avdj}} 
    \label{fig:S_J_kappa_4Mo}
\end{figure*}

In Figures~\ref{fig:S_J_kappa_2Mo} and \ref{fig:S_J_kappa_4Mo} we show snapshots at different times of the supersaturation ratio, $S$, the normalized nucleation rate, $\hat{J}_*$, and the dust opacity, $\kappa_\mathrm{d}$, for the CE simulation with a 1.7~\msun\ and a 3.7~\msun\ AGB star, respectively\footnote{All figure movies can be found at the following URL: \url{https://tinyurl.com/y455avdj}}.  
The quantities $S$, $\hat{J}_*$ and $\kappa_\mathrm{d}$ trace where the dust seeds are more likely to form, where nucleation is actually happening, and where dust has formed, respectively. To aid the visual comparison, all of the snapshots include the same temperature contours, which are equally spaced in linear scale from 2000~K (innermost) to 750~K (outermost). The $x$ and $y$ axes define the binary system's orbital plane while the $z$ axis is perpendicular to that plane.

The dust nucleation rate starts to increase after the supersaturation ratio has exceeded the threshold value $S_c$. As a consequence, the regions with high values of $J_*$ are typically associated with those of high $S$ (Figures~\ref{fig:S_J_kappa_2Mo} and \ref{fig:S_J_kappa_4Mo}).
At early times (3 and 5 years), the efficient formation of dust seeds ($S>10$ and  $\hat{J}_* > 10^{-22}$) takes place in the XY plane in lumpy arcs or crescent shapes around the center of mass.  
In the XZ plane, the regions of seed formation are asymmetric and elongated in the orbital direction, with one side being more efficient than the other. This asymmetry is likely due to the initial mass-loss through the outer Lagrange points, which causes dense spiral outflows where dust can be formed more efficiently.
At later times (after 11.9 yrs), the regions of seed formation tend to be more regular in shape (quasi-spherical) and appear as shells in both models (the lumpy appearance at larger radii is due to low numerical resolution).

Inspecting the $\kappa_\mathrm{d}$ distribution (Figures~\ref{fig:S_J_kappa_2Mo} and \ref{fig:S_J_kappa_4Mo}), dust formation occurs as early as $\sim$1~yr from the onset of simulations, at a distance of $\sim$10~au from the center of mass. However, it is only after $\sim$3 years that a shell of high opacity ($\kappa_\mathrm{d} >$ 5 cm$^2$\,g$^{-1}$) can be seen surrounding the stars. 
For the 1.7~\msun\ simulation, the dusty shell at 11.9 yrs is very thin in some parts and is at $\sim100$ au from the center of mass. With time, its thickness increases (especially at 14.6 and 17.9 yrs) and becomes more uniform. At 20 yrs the dusty shell's radius has increased to $\sim160$ au.
For the 3.7~\msun\ simulation, at 11.9 yrs the dusty shell is also located $\sim100$~au from the center of mass and its thickness increases between 14.6 and 17.9 yrs. However, in contrast to the lower mass simulation, at 20 yrs the shell has a radius of $\sim200$ au. 

In the 1.7~\msun\ simulation the dusty shell is thicker and is elongated in the polar direction, with the development of bulges. In the higher mass model, such polar protuberances are absent and the dusty envelope maintains a more spherical shape.
These differences in the geometry of the dusty shell may be related to the differences in orbital separation and mass ratio between the two models, which affect the mass ejection geometry and thus the dust distribution. 

\begin{figure*}
    \centering
    \includegraphics[width=0.495\linewidth]{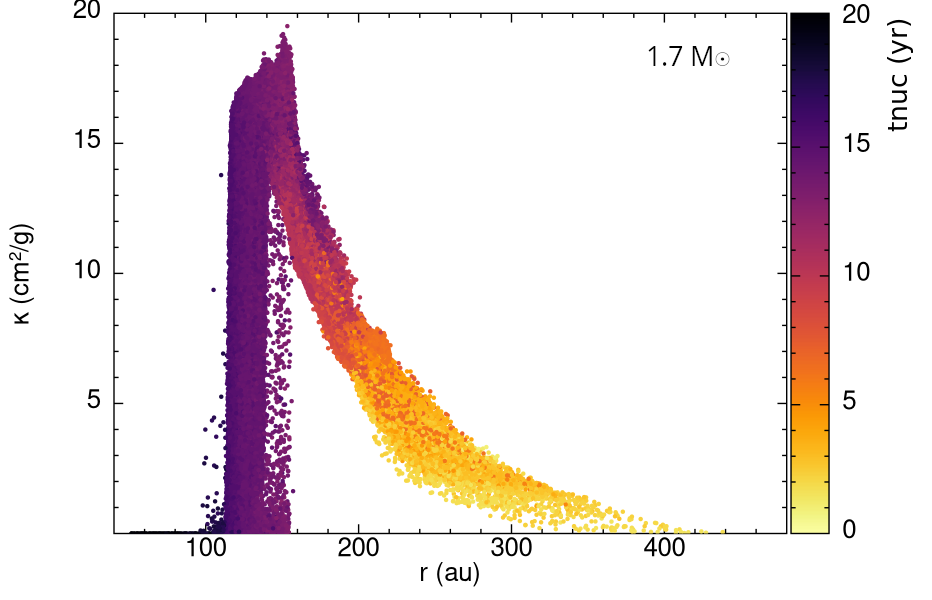}
    \includegraphics[width=0.495\linewidth]{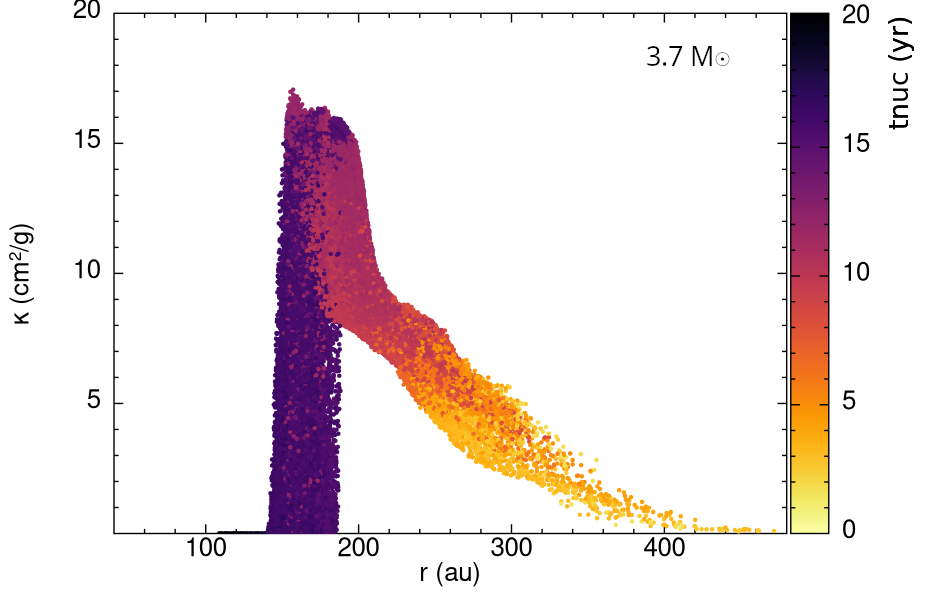}
    \caption{Dust opacity versus distance from the centre of mass for the 1.7~\msun\ (left) and 3.7~\msun\ (right) model at 20.2 years from the start of the simulations. The color map indicates nucleation time, $t_{\rm nuc}$, the time (in years) when dust was formed in the SPH particles, as calculated by the ratio of the normalised moments, ${\widehat{\cal K}_3}/{\widehat{\cal K}_0}$,  which represents the amount of condensed carbon atoms per dust grain. A movie of this figure can be found at the following URL: \url{https://tinyurl.com/y455avdj}}
    \label{fig:opacity-versus-distance}
\end{figure*}

In general, regions with high values of $\kappa_\mathrm{d}$ are located outside the regions with high values of $\hat{J}_*$, and this is because it takes time for the opacity to increase after the gas reaches the condensation temperature. In Figure~\ref{fig:opacity-versus-distance}, the SPH particles are colored according to the time at which dust formation occurs as indicated by the ratio of the normalised moments, ${\widehat{\cal K}_3}/{\widehat{\cal K}_0}$,  which represents the amount of condensed carbon atoms per dust grain.

In both simulations, the opacity peaks at $\sim$17-20 cm$^2$\,g$^{-1}$ (Fig.~\ref{fig:opacity-versus-distance}). However, as the envelope expands, the density and temperature decrease. Dust growth, which occurs through the collision of monomers with dust grains, eventually stops because of the rarefaction of monomers. The number of condensed carbon atoms per gram in the dust, $\mathcal{K}_3/\rho$, then freezes and, according to Eq.~\ref{eq:kappa_d}, the opacity becomes a simple linear function of the temperature and thus decreases as the gas expands and cools.

\begin{figure}
    \centering
    \includegraphics[width=0.426\linewidth]{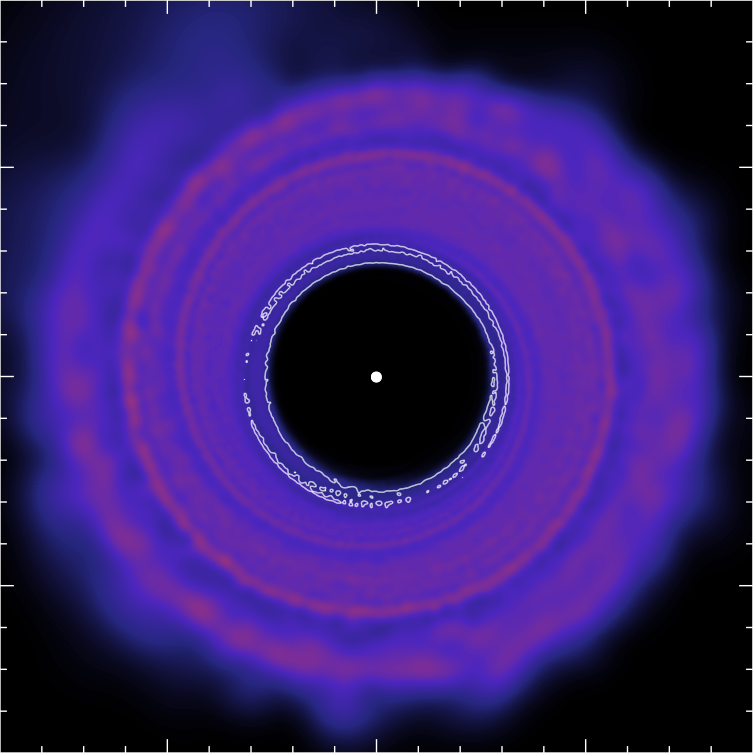}
    \includegraphics[width=0.56\linewidth]{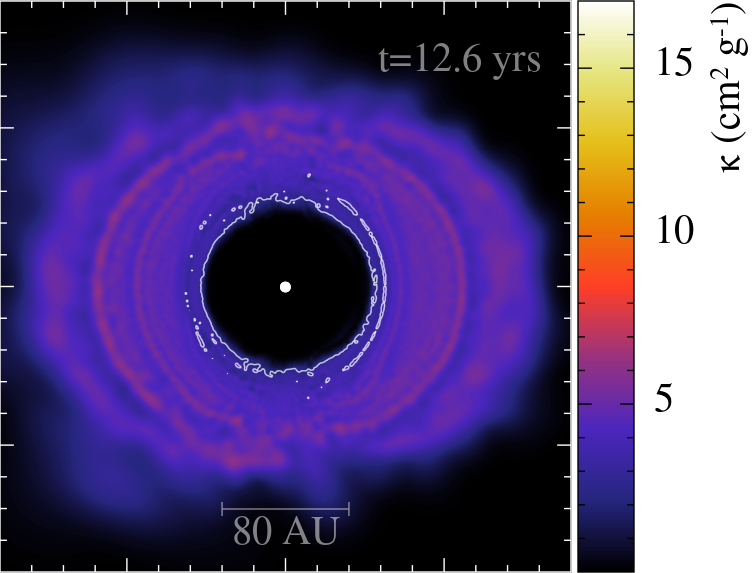}
    \includegraphics[width=0.426\linewidth]{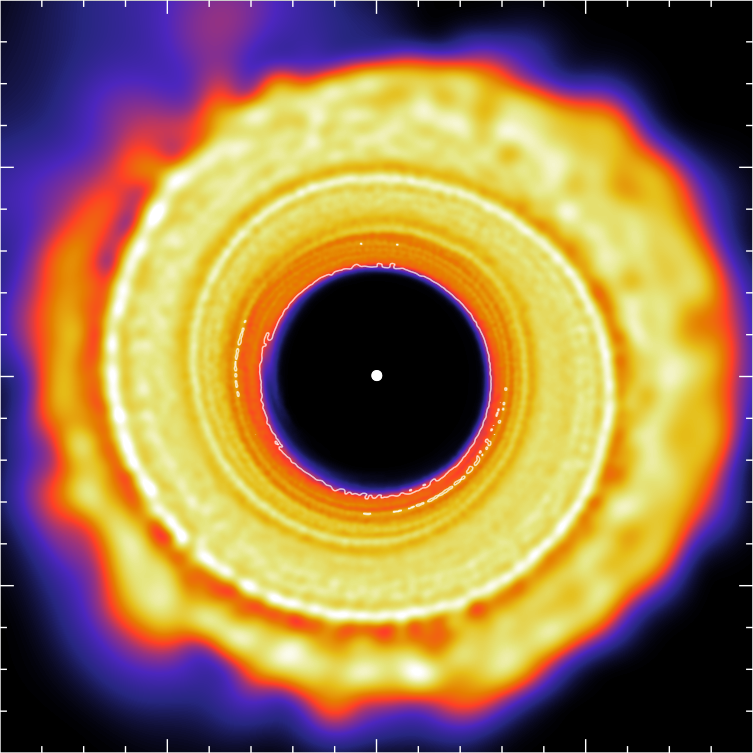}
    \includegraphics[width=0.56\linewidth]{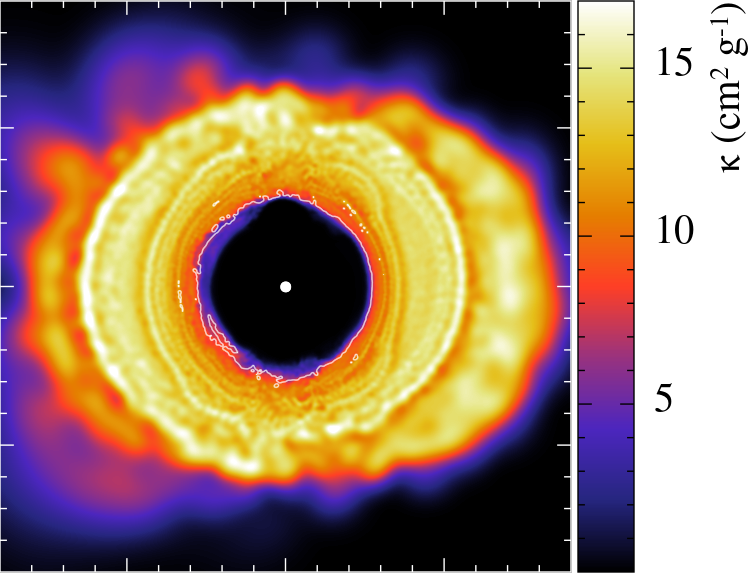}
    \includegraphics[width=0.426\linewidth]{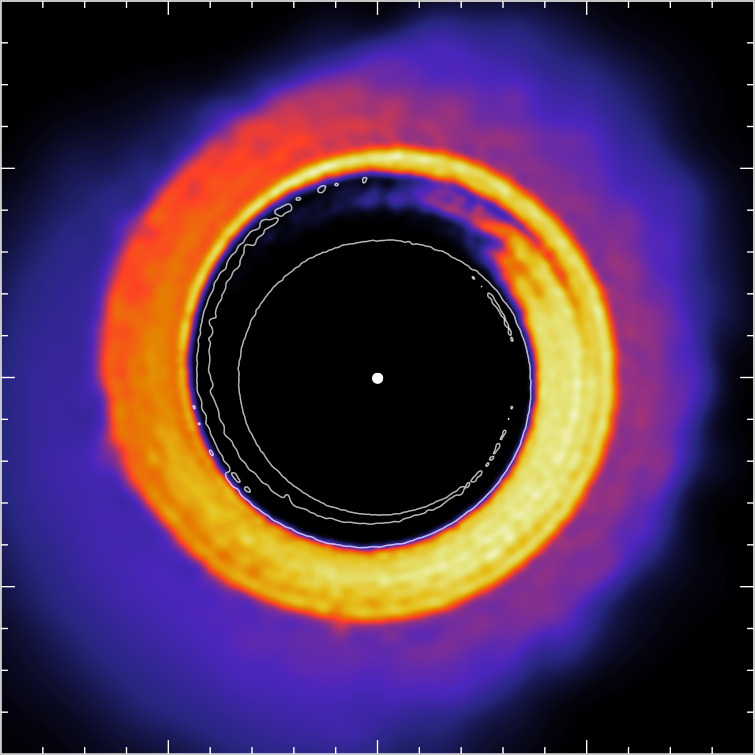}
    \includegraphics[width=0.56\linewidth]{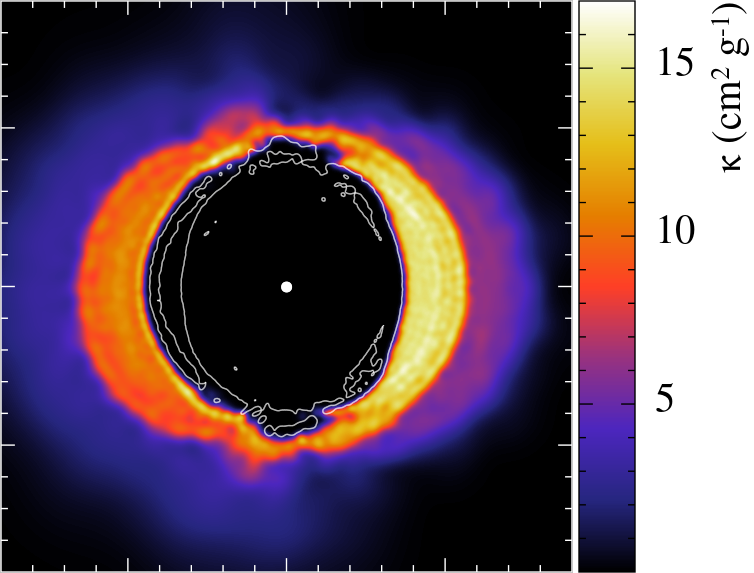}
    \caption{Equatorial (left column) and perpendicular (right column) slices of opacity at 12.6 years for the 1.7~\Msun Bowen simulations \citep{GonzalezBolivar2023} with $\kappa_{\rm max} = 5$~cm$^2$~g$^{-1}$ (top row), with $\kappa_{\rm max}=15$~cm$^2$~g$^{-1}$ (middle row), and for our 1.7~\Msun simulation with dust nucleation (bottom row). The white contour indicates a temperature of 1,500~K. The distance scale is the same for each simulation and is shown in the upper right panel.}    \label{fig:kappas_bowen}
\end{figure}

In the Bowen approximation (\citetalias{GonzalezBolivar2023}), the opacity reaches its maximum, user-defined value, $\kappa_\mathrm{max}$, as soon as the temperature drops below the condensation threshold and remains constant as the envelope expands, which is not the case in our formalism because of the temperature dependence of $Q^\prime_\mathrm{ext}$ in Eq.~\ref{eq:kappa_d}. As a result, using the Bowen approximation leads to more opaque inner regions (as seen in Figure~\ref{fig:kappas_bowen}) and potentially more acceleration by radiation pressure. 

\begin{figure*}
    \centering
    \includegraphics[width=0.45\linewidth]{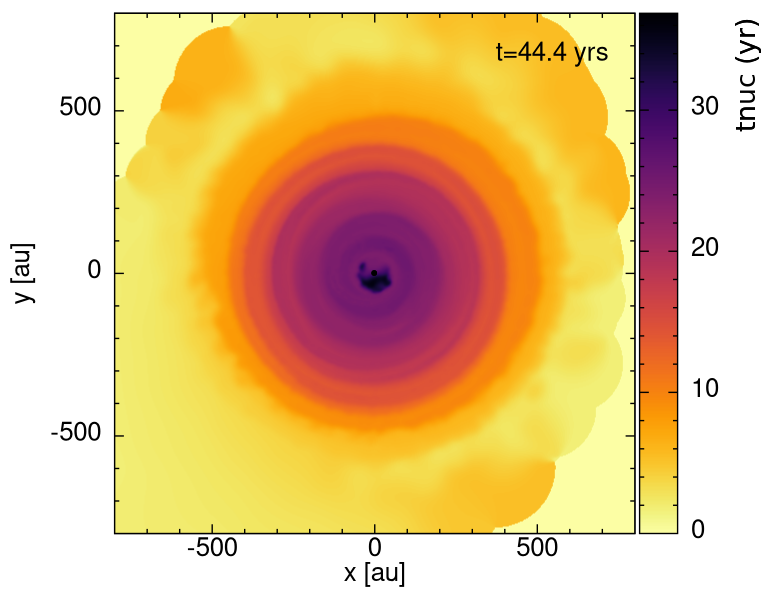}
    \includegraphics[width=0.45\linewidth]{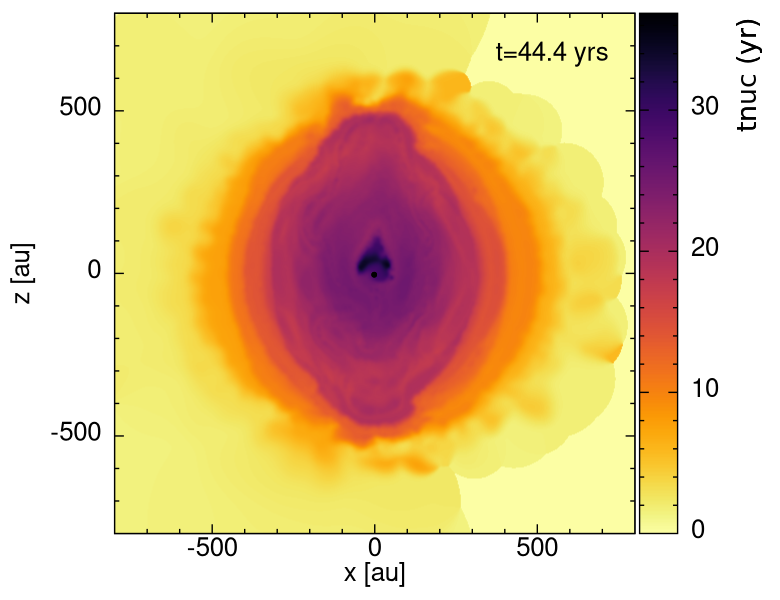}
    \includegraphics[width=0.45\linewidth]{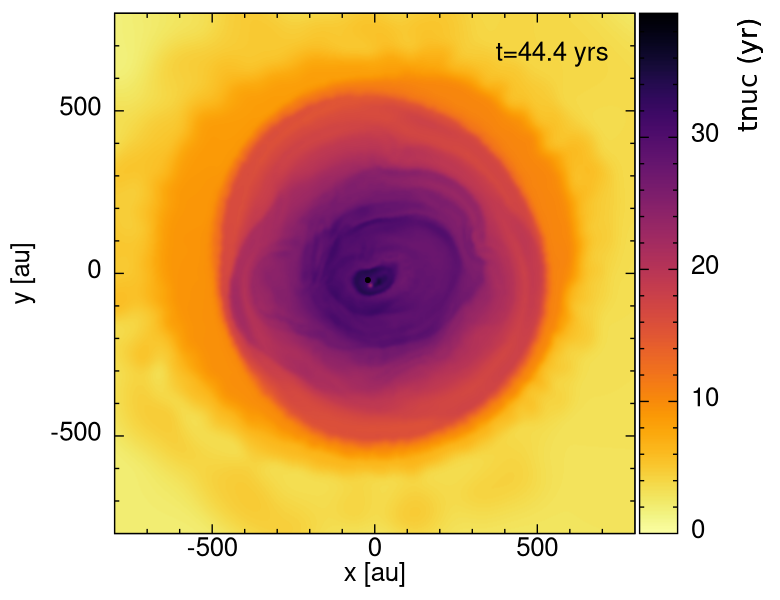}
    \includegraphics[width=0.45\linewidth]{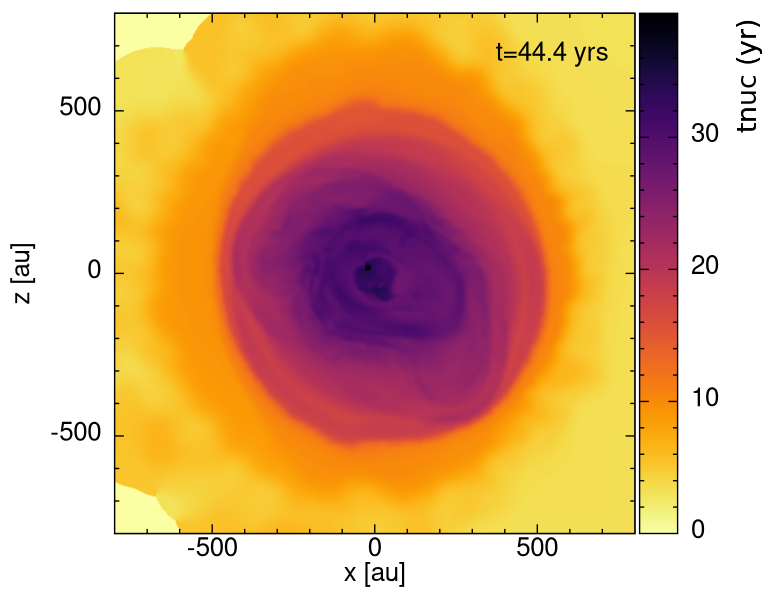}
    \caption{Colour map of nucleation time, $t_{\rm nuc}$, in the XY (left column) and XZ plane (right column), for the 1.7~\msun\ (top row) and 3.7~\msun\ (bottom row) model, after 44.4~yrs of simulation. A movie of this figure can be found at the following URL: \url{https://tinyurl.com/y455avdj}.}
    \label{fig:dustySPH-particles-XY-XZ-2Mo-4Mo}
\end{figure*}

Figure~\ref{fig:dustySPH-particles-XY-XZ-2Mo-4Mo} shows the expansion of the envelope with newly formed dust particles (darker colours) located close to the center and older dust father from the centre (lighter colours). 
The inner region of the 3.7~\Msun{} model shows more irregular structures than the 1.7~\msun\ model, with younger dust overtaking older dust. 
There is no simple relationship that indicates when the dust forms for a given radius because it depends on complex hydrodynamic interactions that occur in the ejected material during the in-spiral phase --- namely, the passage of spiral shocks through the material driven by the binary interaction. A similar result was found by \citet[][see their figure 2]{Iaconi2020}.

\begin{figure*}
    \centering
    \includegraphics[width=0.49\linewidth]{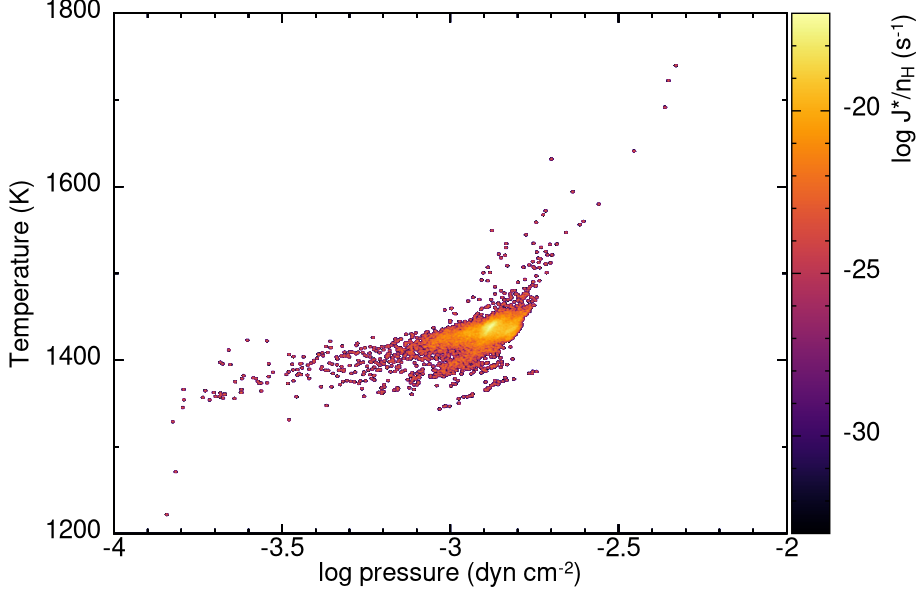}
    \includegraphics[width=0.49\linewidth]{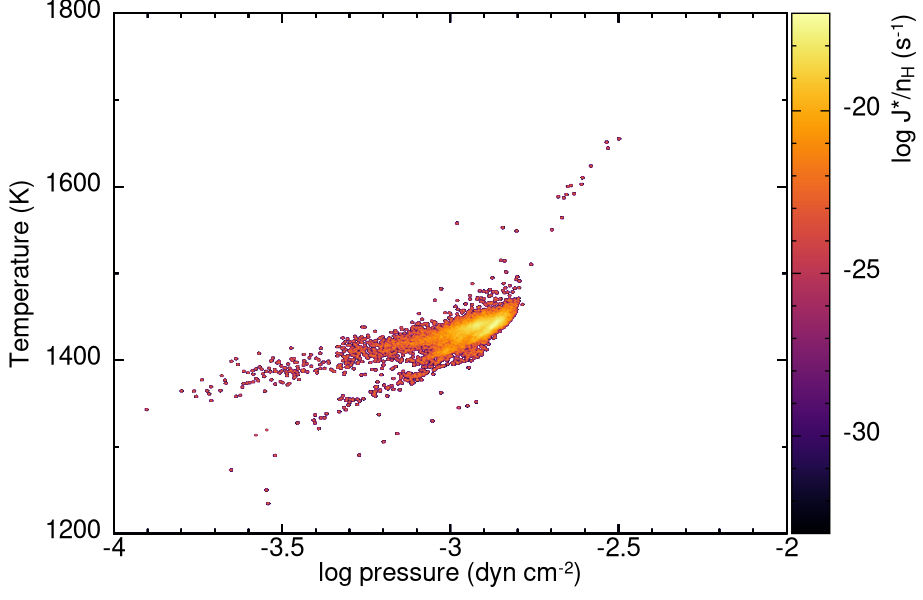}
    \caption{Temperature vs. pressure for SPH particles with dust opacity $\kappa_\mathrm{d}=0.20-0.25$~cm$^2$~g$^{-1}$. The color map is the normalized nucleation rate $J_*/n_{\langle \mathrm{H} \rangle}=\hat{J}_*$, and indicates where dust is forming at the largest rate. The plot is a collection of dusty SPH particles at all times during the CE simulations, from beginning to end.
    The left (right) panel corresponds to the 1.7~\msun\ (3.7~\msun) model.}
    \label{fig:dustySPH-particles-T-vs-pressure}
\end{figure*}

The thermodynamic conditions in the SPH particles at the time the opacity reaches a threshold value of $\kappa_\mathrm{d}=0.02$~cm$^2$~g$^{-1}$ (one hundred times larger than the gas value) are shown in Figure~\ref{fig:dustySPH-particles-T-vs-pressure}. The particles are confined to a narrow region of temperature ($\approx 1450 \pm 100$~K) and  pressure ($\log_{10} P \approx -2.9 \pm 0.1$ dyn\,cm$^{-2}$), where the nucleation rate is the highest.

\subsubsection{Dust size}
\label{ssec:dust_size}

\begin{figure}
    \centering
    \includegraphics[width=\linewidth]{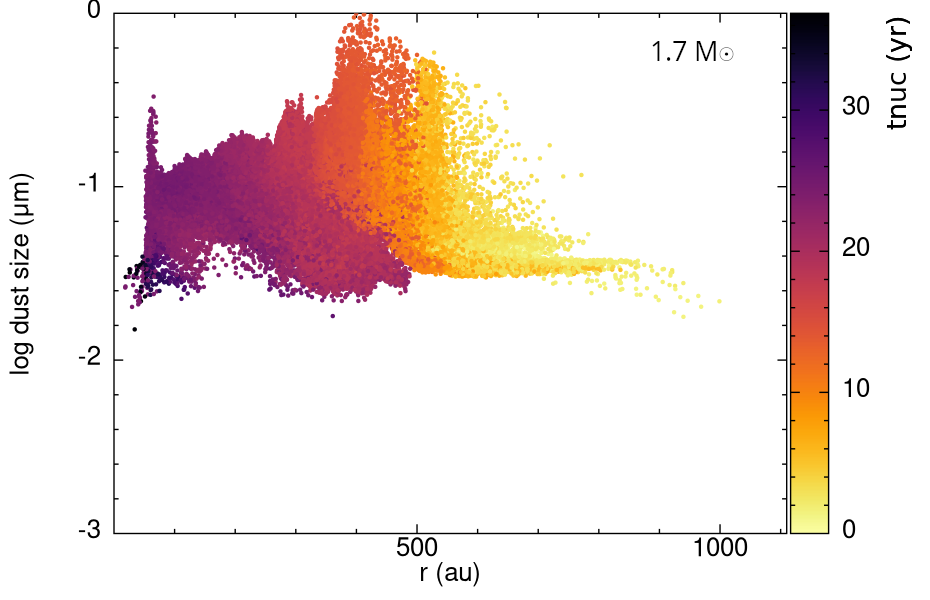}
    \includegraphics[width=\linewidth]{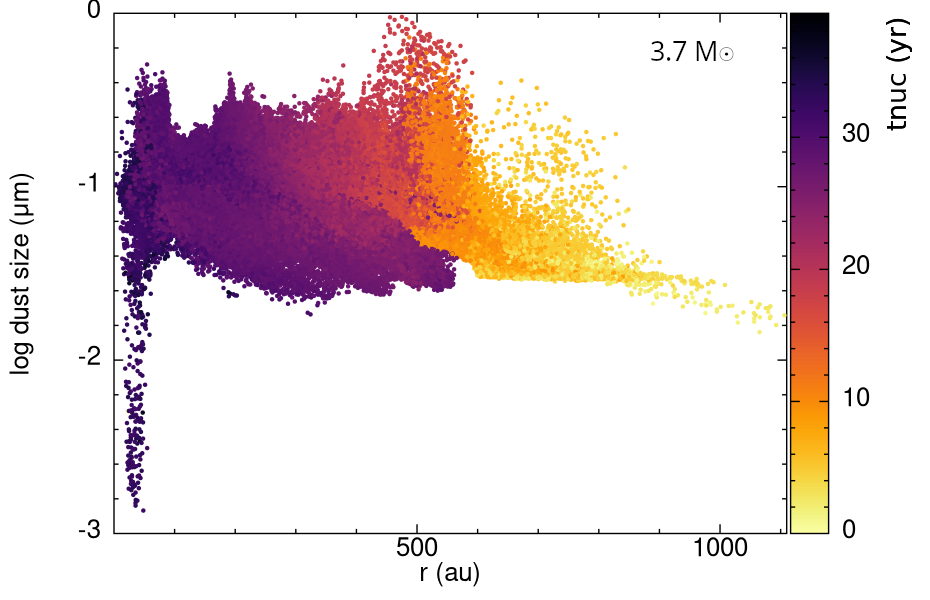}
    \caption{Average dust grains size versus distance from the centre of mass, for the 1.7~\msun\ (top) and 3.7~\msun\ (bottom) models, after 44.4 years of simulation. The color map indicates the nucleation time in years. A movie of this figure can be found at the following URL: \url{https://tinyurl.com/y455avdj}} 
    \label{fig:median-size-distribution}
\end{figure}

Figure~\ref{fig:median-size-distribution} shows the mean size of dust grains for both simulations in the SPH particles as a function of radius, at an age of $\sim$44~years. The different colors indicate the time with respect to the start of the simulations at which dust formation happened, with younger dust in darker colours. For SPH particles with $t_\nuc \lesssim 30$~yr, the grains have a minimum size of $\sim 0.03 \mu$m. The upper limit depends on the time of formation: early dust (yellow colour) forms grains in a narrow range of sizes (0.03 - 0.04~$\mu$m), while at later times, around $t_\nuc \sim$12-15~yrs, it can reach a maximum size of 1~$\mu$m. The main difference in Figure~\ref{fig:median-size-distribution} between the 1.7 and 3.7~\Msun{} models, is the persistence, in the more massive AGB star, of active dust nucleation in a shell around 10-20~au. This is attested by the co-existence in this region of newly formed small grains $10^{-3}$~$\mu$m with bigger ones that can reach up to $\sim 0.4 \mu$m.

It should be noted that in the absence of dust destruction in our formalism, the average grain size of an SPH particle can still decrease, if newly formed, small grains are being produced. This explains in particular why the average grain size distribution of Figure~\ref{fig:median-size-distribution} is not increasing with time.

\begin{figure*}
    \centering
    \includegraphics[width=0.49\linewidth]{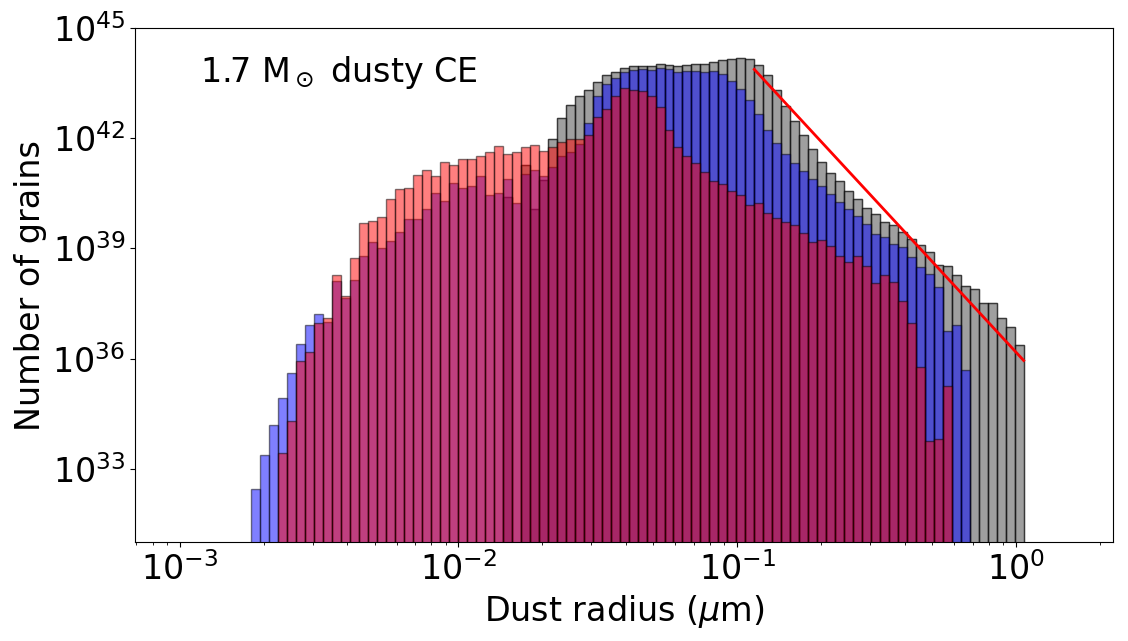} 
    \includegraphics[width=0.49\linewidth]{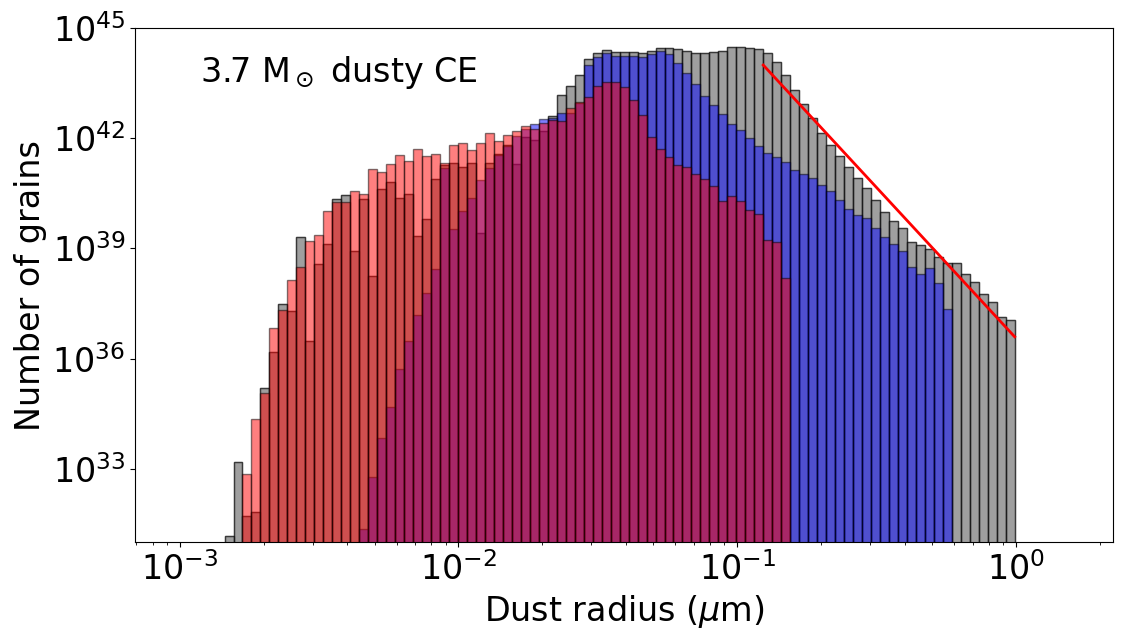} 
    \caption{Histograms of the {\it average dust grain size in each SPH particle, not the individual grain size}. The pink, blue and gray histograms are at $t=7.0$~yr, $t=20$~yr and $t=44$~yr from the start of each simulation, respectively. We fit part of the gray histogram data using a power law, depicted here using a red line. We note that this may not correspond to the grain size distribution and cannot therefore be directly compared to, for example, MRN \citep{Mathis1977}.
    }
    \label{fig:histogram median-size}
\end{figure*}

Figure~\ref{fig:histogram median-size} displays histograms of the average dust grain size of all SPH particles, at three different times. For the 1.7~\msun\ model the range of dust sizes at 7 and 20~yrs remains approximately the same, between 0.002 - 0.6~$\mu$m, but by 40 years, it has grown to 0.02 - 1~$\mu$m. The evolution of the average dust grain size in the  3.7~\msun\ model is quite different: at 7~yrs dust grains are confined to a narrow range (0.002 - 0.15~$\mu$m) increasing to 0.004-0.7~$\mu$m at 20~yr, and then increasing again to 0.002-1~$\mu$m by the end of the simulation at 44~yr. 
The peak of the dust distribution is very similar in the two models: 0.04~$\mu$m at 7 years, 0.05 - 0.07~$\mu$m at 20 years and 0.05 - 0.2~$\mu$m at 44 years.

At the end of the simulations, the largest average grain sizes are approximately 0.8-1.0~$\mu$m. The distributions are characterized by an initial positive slope, with the number of grains increasing with grain size from $\approx 0.001$ up to $\approx 0.05$~$\mu$m, then a plateau up to $\sim$0.1 and 0.15~$\mu$m for the 1.7~\msun\ and 3.7~\msun\  models, respectively, and finally a decrease as a function of the average grain size. We fit the distribution of the large grains with a power law of the form:
\begin{equation}
    n(a_{\rm ave}) = C a_{\rm ave}^{-q}
\end{equation} 
where $a_{\rm ave}$ is the average grain size in an SPH particle in $\mu$m. We find $q=8.2$ for both models by fitting the slope by eye
(red line in Fig.~\ref{fig:histogram median-size}) with $C=1.5\times10^{36}$ for the 1.7~\msun\ model and $C=3.6\times10^{36}$ for the 3.7~\msun\ model.
In their study of dust formation in CE, \citet{Iaconi2020} used a post-processing code and found that the grain size distribution of their carbon dust  always had a negative, albeit changing, slope. In their models, the largest grains have a slope of $q=6.6$, only slightly smaller than our  value. 
The difference is not so large considering that (i) the donor star used by \citet{Iaconi2020} is an RGB star of 0.88~\Msun{} and 83~\Rsun, (ii) the structure of the outflow and thus the conditions for dust growth are likely to be impacted by the differences in the initial conditions, (iii) the dust formation model is based on a different formalism  \citep{Nozawa2013} that, contrary to ours, does not assume chemical equilibrium.

While the dust size distribution is only the distribution of the average grain size of SPH particles, it is notable that the steepness of the slope is much larger than the interstellar dust grains value given by the model of \cite{Mathis1977}. 

\subsubsection{Dust mass}
\label{ssec:dust_mass}

\begin{figure}
    \centering
    \includegraphics[width=1.0\linewidth]{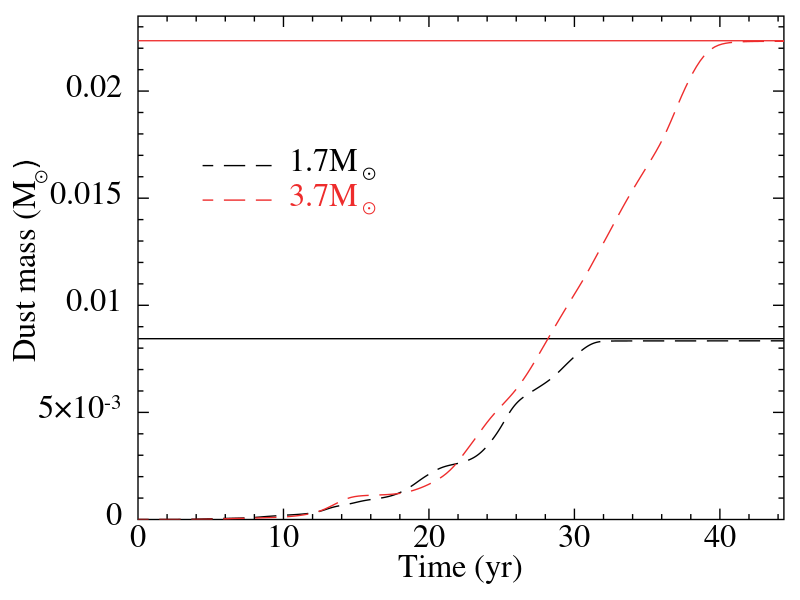}
    \caption{Total dust mass as a function of time for the 1.7~\Msun\ (black dashed line) and 3.7~\Msun\ (red dashed line) models. The horizontal lines indicate the maximum amount of carbon available in the envelope of the 1.7~\Msun\ (black solid) and 3.7~\Msun\ (red solid) models (Eq.~\ref{eq:maximum_carbon}).} 
    \label{fig:dust-mass-vs-time}
\end{figure}

Figure~\ref{fig:dust-mass-vs-time} shows the evolution of the total dust mass (Eq.~\ref{eq:dust_mass}) for the two simulations. It takes about 10~yr for the dust production to really take off, after which the dust mass increases steadily until it reaches a plateau. For the 1.7~\msun\ (3.7~\msun) model this plateau is reached at $\sim$32~yr (38~yr) for a value of $M_\dust  \approx8.4\times10^{-3}$~\msun ($2.2\times10^{-2}$~\msun).
The rapid increase in dust production is a consequence of a large fraction of the expanding envelope reaching temperatures below the condensation threshold (see Sect.~\ref{sec:photo}). When nucleation is efficient (i.e., when the number of small dust grains is large), a lot of monomers are available and contribute to dust growth, producing the increase in the dust mass and the spreading in the grain size distribution seen in Figure~\ref{fig:histogram median-size}. 
With the expansion of the envelope, the nucleation rate decreases, fewer monomers  form and dust production reduces until it  stops when all the gas has been ejected. We checked that the dust mass produced by the models does not depend on the numerical resolution (see Appendix~\ref{AppendixB}).

Comparing the dust production in our 2 simulations indicates that the more massive progenitor produces dust for a longer period of time (40 vs 30~yr) and makes three times more dust than the 1.7~\msun model, but this scales with the envelope mass which is about 3 times more massive in the 3.7~\Msun{} model. So it seems that dust formation is equally efficient in our models. Figure~\ref{fig:dust-mass-vs-time} indeed indicates that dust production, estimated as the ratio of the dust mass to the available mass of carbon in the envelope, is of the order of 100~per~cent for both models.
Caution should be exercised with regard to dust formation efficiency, as no dust destruction mechanism is considered here, so such efficiency should only be considered as an upper limit. How this percentage is reduced when dust grain destruction processes are included will be covered in further work.

\citet{Iaconi2020} also described the evolution of the dust mass. In their simulation the total dust mass plateaus at $\sim2.2\times10^{-3}$~\msun\ after $\sim$14 years for a 0.88~\Msun\ RGB star, which started off with a radius of 83~\Rsun. 
The amount of dust produced in our 1.7~\Msun\ simulation is four times larger for a star twice as massive. 

\subsection{Orbital evolution, unbound mass and comparison with non-dusty simulations}
\label{ssec:orbital-evolution}

For all simulations, the dynamic in-spiral phase ends by $\sim$7.5 yr (Figure~\ref{fig:separation_massunbound}, upper and lower left panels). For the ``high opacity'' simulations (i.e. including nucleation or the high Bowen $\kappa_\mathrm{max}$ value) the in-spiral is somewhat faster than non-dusty simulations or those with low Bowen maximum opacity. 
For the 1.7~\msun\ models the final separation is $\sim$30~per~cent larger (40-42~\Rsun) for these high opacity runs compared to the other simulations (33~\Rsun\  and somewhat still decreasing). On the other hand, for the 3.7~\msun\ models the final separations are very similar for all models ($10-12$~\Rsun) and very close to the core softening radius\footnote{The softening radius introduces a smoothing function to the gravitational potential of the point masses, reducing its strength at short distances. It is used to prevent unrealistic high forces between SPH particles and the point masses and to improve numerical stability in simulations.} (8~\Rsun), which tends to prevent the binary cores from getting closer. Note that the 3.7~\Msun\ models have a relatively low $q=M_2/M_1=0.16$, but the companion is unlikely to merge with the primary's core.

To determine the ejected mass we define the ``mechanical'' bound mass, whose sum of kinetic and gravitational potential energy is negative ($E_\mathrm{tot}^\mathrm{m}=E_\mathrm{kin}+E_\mathrm{gr}<0$), and the thermal bound mass, for which the sum of kinetic, gravitational {\it and thermal} energies is less than zero ($E_\mathrm{tot}^\mathrm{th}=E_\mathrm{kin}+E_\mathrm{gr}+E_\mathrm{th}<0$, where $E_\mathrm{th}$ only includes the thermal component, but excludes the recombination energy\footnote{Including recombination energy would assume that the entire envelope recombines and that all this energy can be transferred to the gas which may not be the case as some may be radiated away. Since the efficiency of recombination energy is still debated, we prefer to use this stricter criterion for the definition of the thermal unbound mass.}).
The amount of bound mass (Figure~\ref{fig:separation_massunbound}, right panels), decreases at a similar rate during the first 12.5~years for all simulations, regardless of whether dust opacity is included or not. This is also indicated in the eight column of Table~\ref{tab:simulations}, which shows a similar percentage of unbound mass for all simulations. 
Note that not all of the envelope is unbound by 12.5 years in the non-dusty simulation. On the other hand the simulation with dust nucleation (blue lines in Figure~\ref{fig:separation_massunbound}), which ran for longer, shows that almost the entire envelope is unbound after 20 years from the beginning of the CE.

\begin{figure}
    \centering    
    \includegraphics[width=\linewidth]{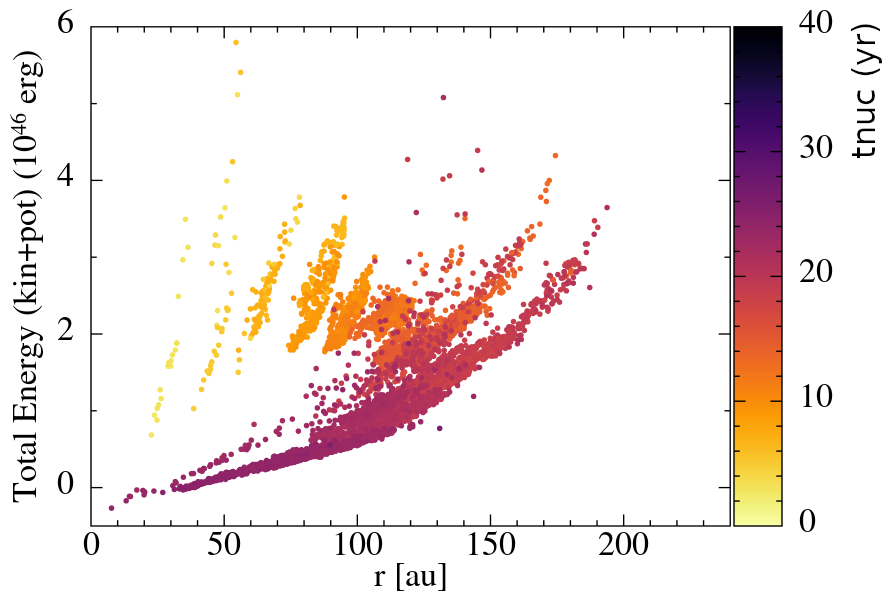}
    \includegraphics[width=\linewidth]{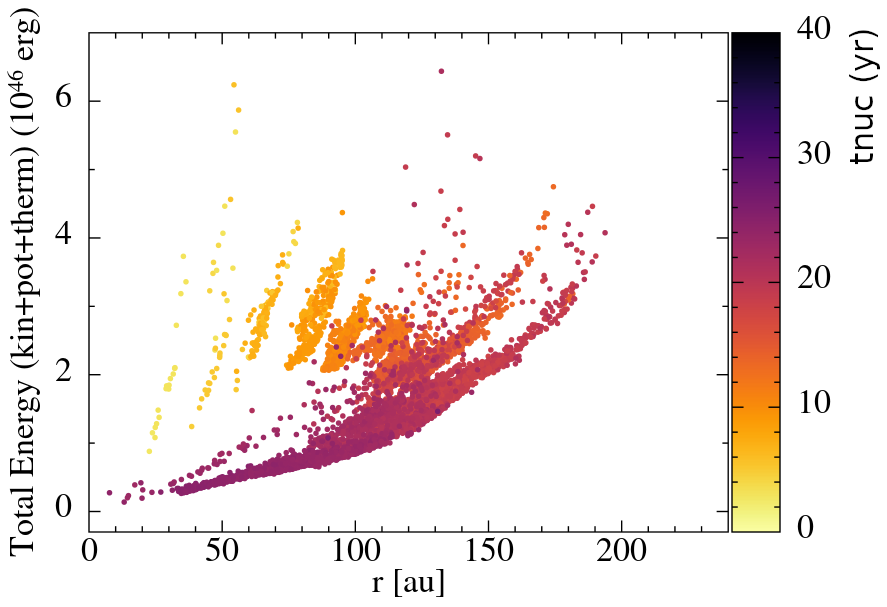}
    \caption{Total energy versus distance from the centre of mass for SPH particles in the 1.7\Msun{} model with dust opacity $\kappa_\mathrm{d}=0.020-0.025$~cm$^2$~g$^{-1}$. The upper panel does not consider thermal energy, while the lower panel does. The color map indicates nucleation time, $\rm t_{nuc}$, as in Figure~\ref{fig:opacity-versus-distance}.
    Similar results are obtained for the 3.7~\Msun{} model. 
    }
    \label{fig:Etot_dust}
\end{figure}

The similarity of the bound mass curves between dusty and non-dusty models suggests that both simulations would unbind the entire envelope, leading to the conclusion that dust has little impact on unbinding the CE mass in these simulations. This claim is supported by Fig.~\ref{fig:Etot_dust} that shows that the total energy of the SPH particles with newly formed dust in the 1.7~\Msun{} model is positive (more so if we include thermal energy in the definition of total energy). The same results are obtained for the 3.7~\Msun{} model. In these simulations dust forms in already-unbound material, which explains why dusty simulations do not unbind more gas than the non-dusty simulations. 

\subsection{The photospheric size}
\label{sec:photo}

In this Section we analyse the size of the photosphere, noting that after dust forms in the neutral regions of the expanding envelope, it provides sufficient opacity to effectively become optically thick. It is thus the expanding dust shell that is likely to be seen from the outside.

\begin{figure*}
    \centering
    \includegraphics[width=0.46\linewidth, trim={0 90 0 70}, clip]{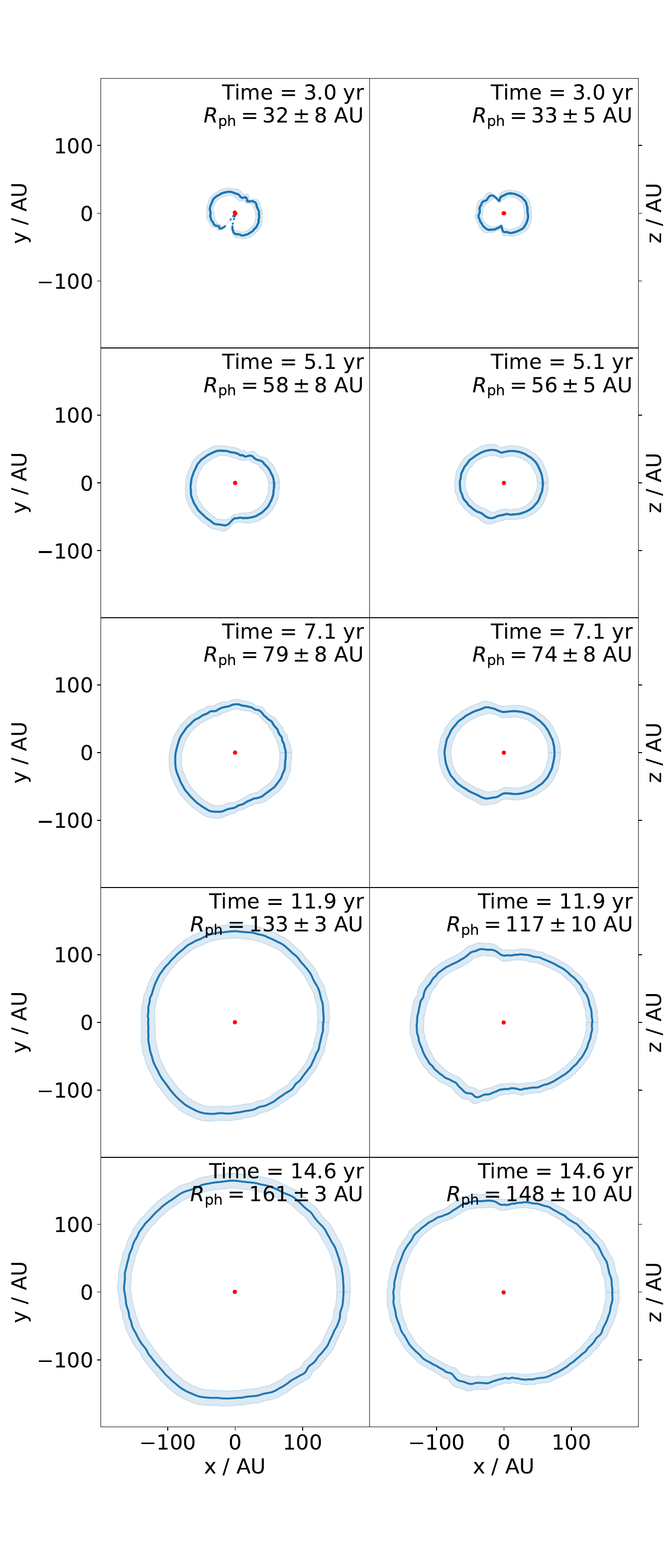}
    \includegraphics[width=0.46\linewidth, trim={0 90 0 70}, clip]{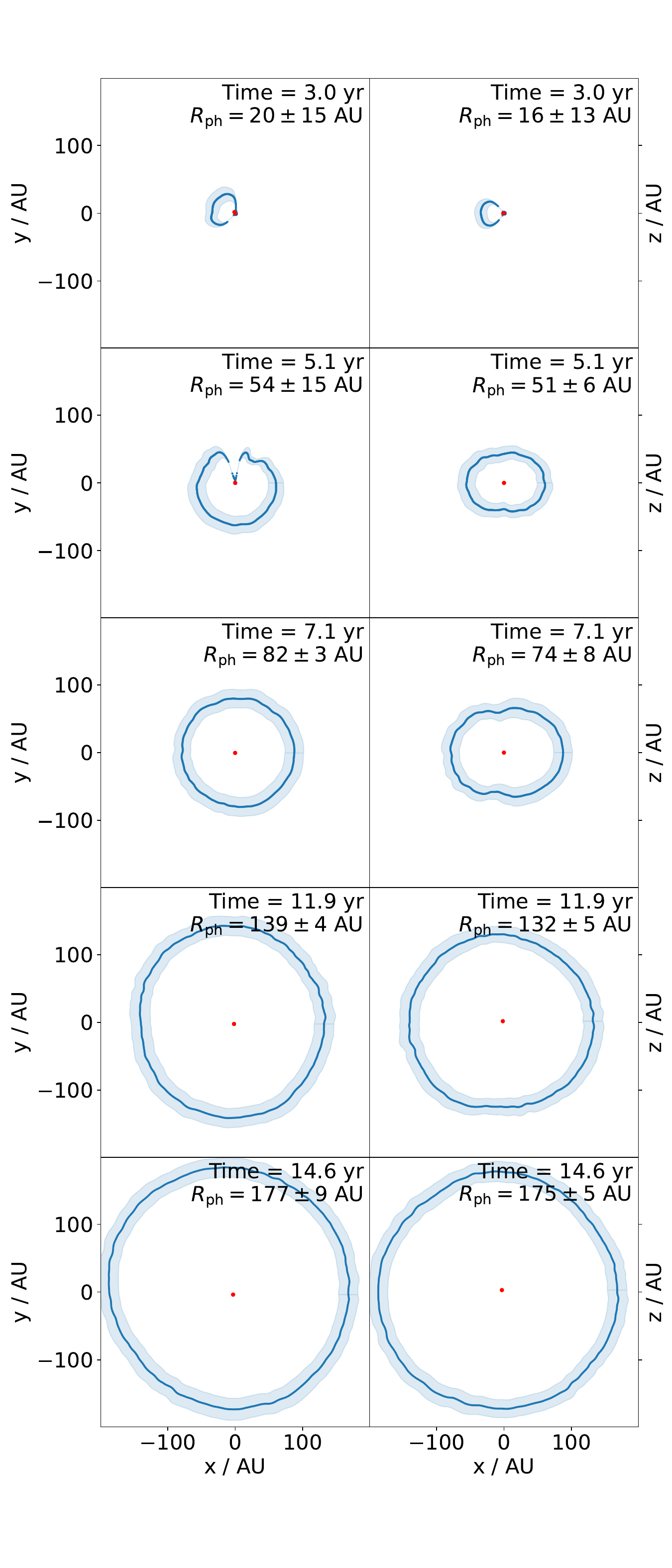}
    \includegraphics[width=0.46\linewidth, trim={0 35 0 0}, clip]{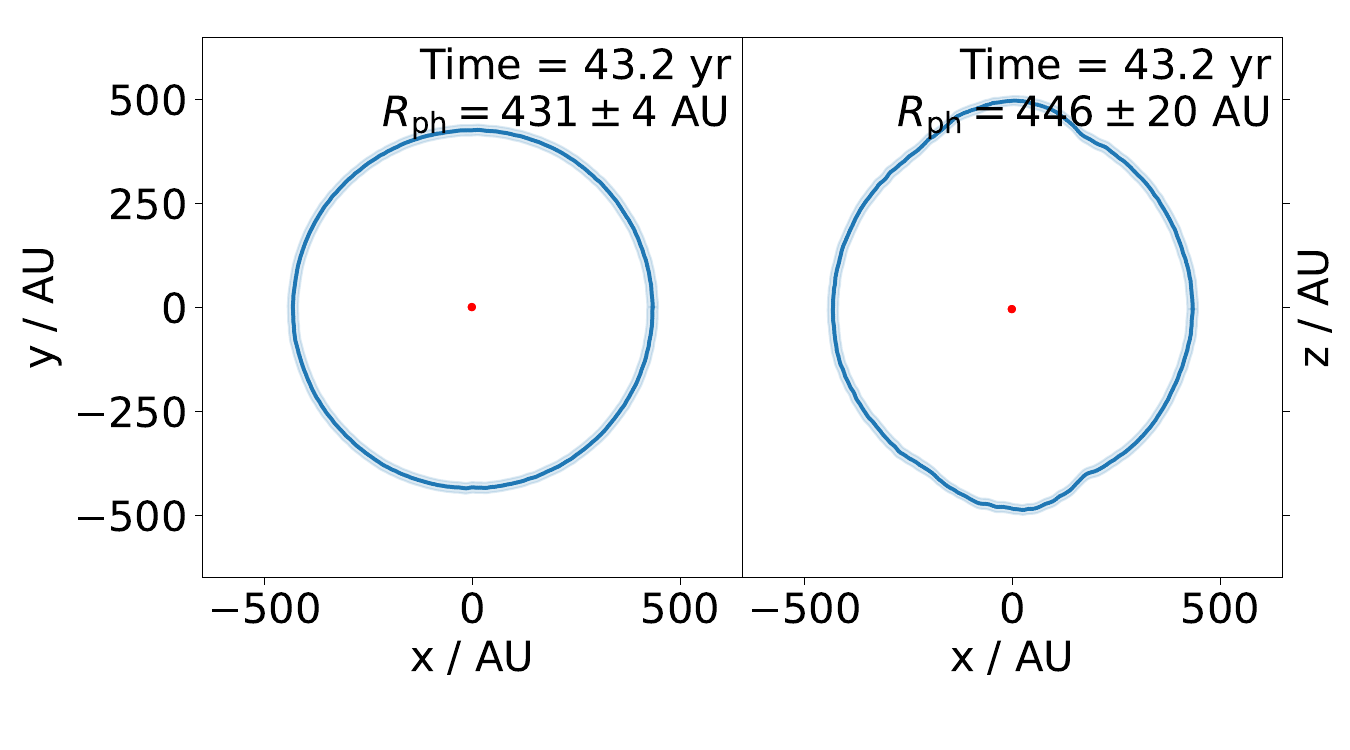}
    \includegraphics[width=0.46\linewidth, trim={0 35 0 0}, clip]{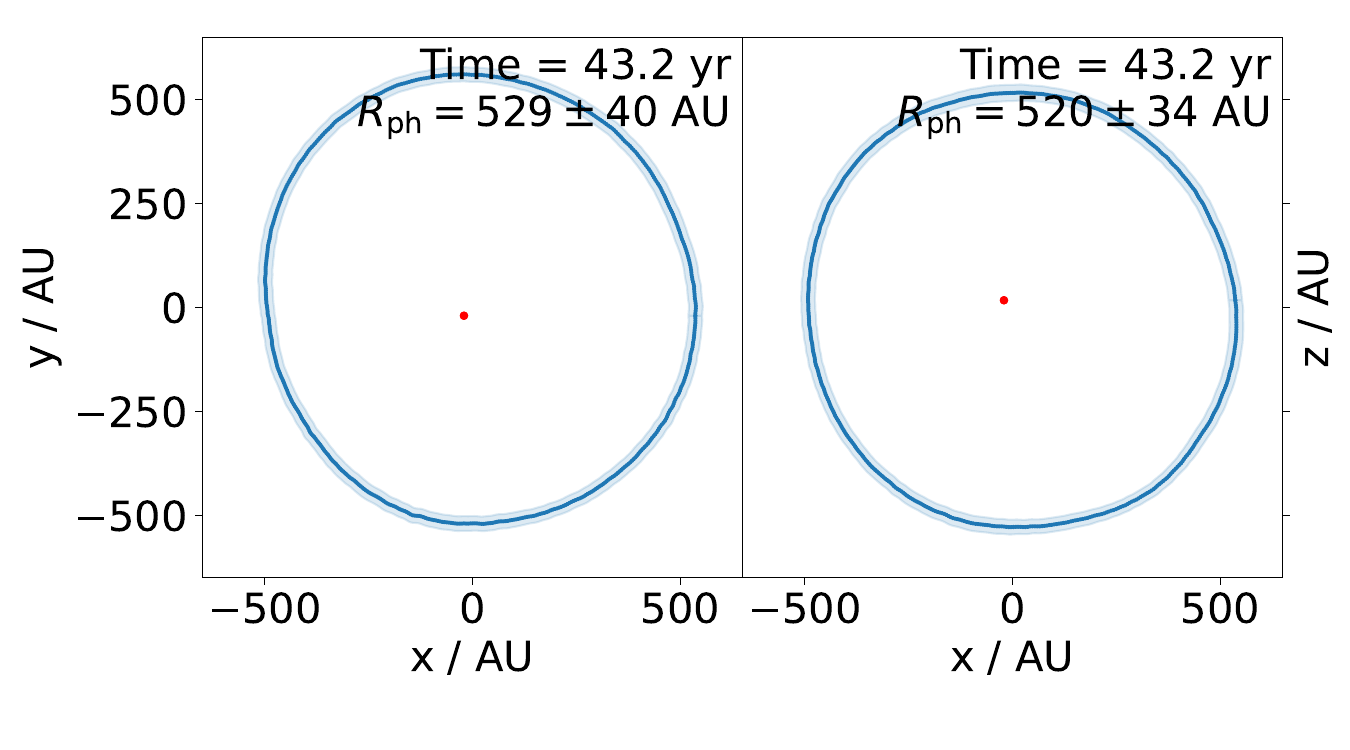}
    \caption{
    Photosphere cross section in the XY (1st and 3rd columns)
    and XZ plane (2nd and 4th columns) for 1.7~\msun\ (left 2 columns) and  3.7~\msun\ (right 2 columns) simulations.
    Blue lines mark the photosphere, the shaded area corresponds to the local smoothing length $h$, 
    and the red dots are the point mass particles.
    The cross section is obtained by
    tracing 600 rays, coming out of the primary star and heading towards all directions in the observation plane, and locating the intersections (marked as blue dots) between these rays and the photosphere (i.e., where the optical depth $\tau$ reaches 1).
    The root mean square of the distance between these intersections and the primary star is marked as the photosphere radius, $R_{\rm ph}$.
    Note that the error on $R_{\rm ph}$ is the standard deviation of $R_{\rm ph}$ for each ray, and it is not the root mean square of $h$.
    }
    \label{fig:photo-xsec}
\end{figure*}

\begin{figure*}
    \centering
    \includegraphics[width=0.49\linewidth]{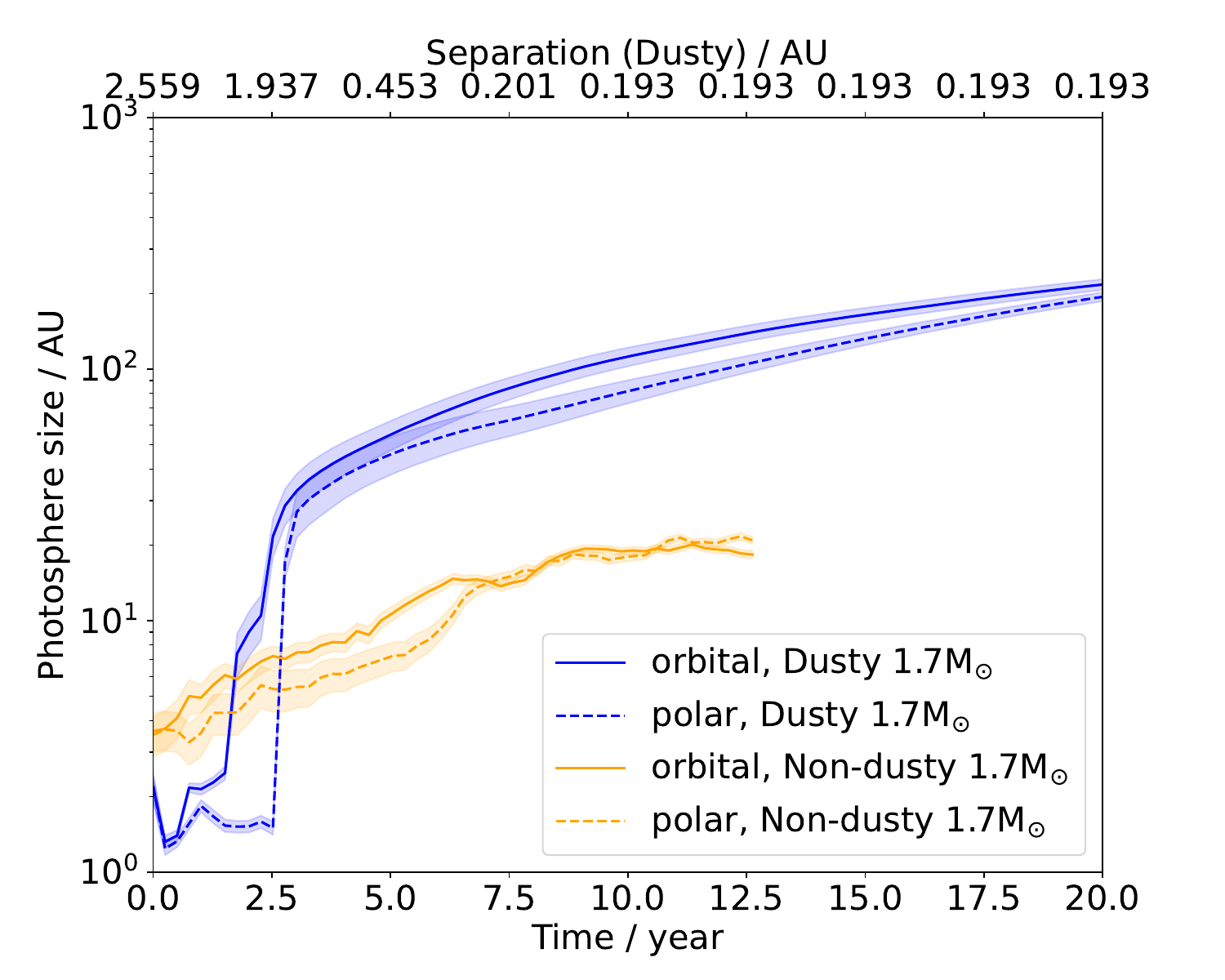}
    \includegraphics[width=0.49\linewidth]{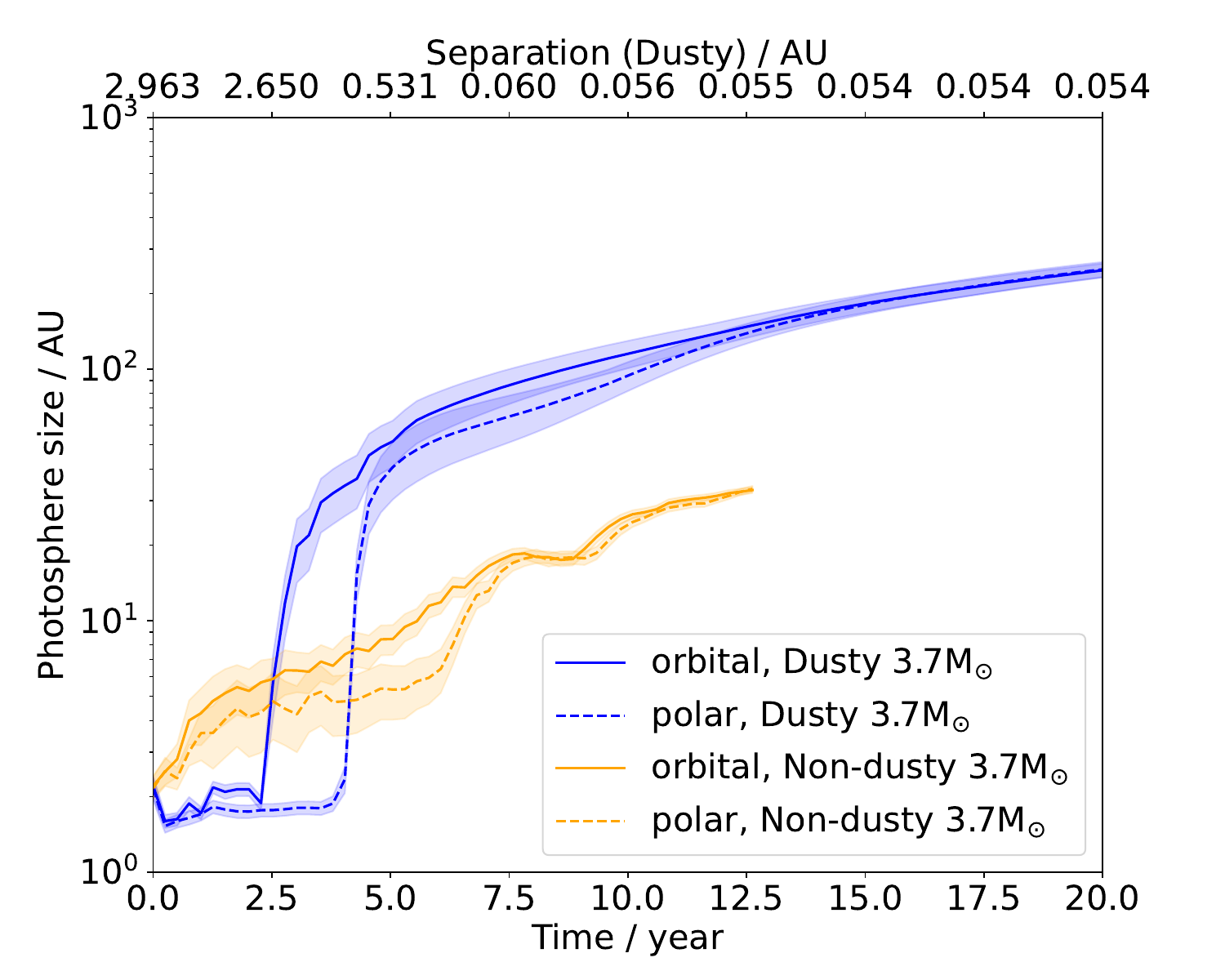}
    \caption{
    Photosphere size evolution in the orbital plane (solid line) and in the polar direction (dashed line) for the 1.7~\Msun\ (left panel) and the 3.7~\Msun{}  (right panel)  simulations with dust nucleation (blue lines) and without dust (orange lines).
    The orbital separation (for the simulations with dust) is marked on the top x-axis.
    The shaded area are the root mean square of the smoothing length interpolated at the intersection between the photosphere and the aforementioned directions, respectively.
    }
    \label{fig:photo-size-vs-sep}
\end{figure*}

Figure~\ref{fig:photo-xsec} shows the photospheric radius along the orbital and perpendicular planes for different moments in time. The radius is obtained by drawing 600 rays emanating from the AGB core point mass particle. The rays are evenly spaced from each other at equal angular intervals.
The optical depth, $\tau$, is obtained at all locations on the rays, integrating inwards to give
\begin{equation}
    \tau(\mathbf{r}) = \int^\infty_r \kappa \rho \, d\mathbf{l},
\end{equation}
where $\kappa$ is the opacity, $\rho$ is the density and the integration is over the ray of direction $d\mathbf{l}$. The integrand in the previous expression at a given position along the ray is computed using

\begin{equation}
    \kappa \rho (\mathbf{r}) = \sum_j m_j \kappa_j W(|\mathbf{r} - \mathbf{r}_j|, h_j),
\end{equation}
where $m_j$ is the (constant) mass of particle $j$ and $W$ the smoothing kernel, which is a function of the position vector, $\mathbf{r}_j$, and of the particle's smoothing length $h_j$. Thus
\begin{equation}
    \tau = \sum_j \frac{\kappa_j m_j} {h_j^2} w_{\mathrm{col}}(q_j),
\end{equation}
where, $\kappa_j$ is the particle's opacity and $q_j$ is the dimensionless distance of the particle to the ray in units of $h_j$.
The dimensionless column kernel, $w_{\mathrm{col}}(q_j)$, is obtained by integrating the smoothing kernel $W(|\mathbf{r} - \mathbf{r}_j|, h_j)$ along $d\mathbf{l}$
\begin{equation}
    w_{\mathrm{col}}(q(x, y, z, h))
    = h^2 \int W(\mathbf{r}(x, y, z), h) d\mathbf{l} .
\end{equation}
The actual direction $d\mathbf{l}$ is not important since the smoothing kernel, $W$, is selected to be spherically symmetric. We use the cubic M$_4$ B-spline kernel from \cite{Schoenberg1946} (see \citet{Monaghan1992}) as the smoothing kernel.

The particle's opacity in the nucleation simulations is the sum of a constant gas opacity, $\kappa_{\gas} = 2 \times 10^{-4} \mathrm{cm}^2 \mathrm{g}^{-1}$, and of the dust opacity, $\kappa_\mathrm{d}$, (as described by Eq.~\ref{eq:kappa_d}), giving
\begin{equation}
    \kappa_j = \kappa_{\gas} + \kappa_{\mathrm{d}, j} .
\end{equation}
In the non-dusty simulations (Figure~\ref{fig:photo-size-vs-sep}), $\kappa_j = \kappa_\gas$ when the temperature $T < 6000$~K, and $\kappa_j = 0.2(1+X) = 0.34 \ \mathrm{cm}^2 \mathrm{g}^{-1}$ otherwise (assuming a hydrogen mass fraction $X = 0.7$).
Hence, the optical depth contribution from particle $j$, $d\tau_j$, is
\begin{equation}
    d\tau_j = \frac{\kappa_j m_j} {h_j^2} w_{\mathrm{col}}(q_j).
\end{equation}
For each ray, particles that are within $2h_j$ from the ray where $w_{\mathrm{col}}(q_j)$ is non-zero, are selected and arranged from closest to the observer to farthest, and the optical depth at any given particle's projection on the ray (which is a line here) is then determined by summing the $d\tau$ from the observer to that particle $n$ via 
\begin{equation}
    \tau = \sum_{j=0}^{n} d\tau_j .
\end{equation}
The photosphere is then found by interpolating the array of $\tau$ (at each particles' projected location on the ray) to find the location where $\tau = 1$. This is the location of the photosphere, marked as a blue line in Figure~\ref{fig:photo-xsec}. At earlier times, the photosphere does not appear to enclose the central binary because dust formation has just started and the high opacity regions may only exist on one side of the binary. The absence of a dusty photosphere simply implies that, at those times, part of the photosphere is much closer to the binary, and the main source of opacity is the gas opacity at the boundary between ionised and recombined gas.

The smoothing length, $h$, at $\tau = 1$ is represented by the light blue region in Figure~\ref{fig:photo-xsec} and gives an idea of the uncertainty on the photosphere's location. It is calculated by using the local density to obtain the value of $h$, the smoothing length  of particles local to the photosphere, from the  relation 
\begin{equation}
    h = h_{\mathrm{fact}} \left( \frac{m}{\rho} \right)^{1/3},
\end{equation}
\citep{Price2018}, where the dimensionless factor $h_{\mathrm{fact}} = 1.2$, and the particle mass, $m$, is a constant for a given simulation.

\begin{table*}
    \centering
    \begin{tabular}{ccccccc}
    \hline
    Time & $R_{\rm phot,1.7}$ & $R_{\rm phot,3.7}$ & $T_{\rm phot,1.7}$ & $T_{\rm phot,3.7}$ & $v_{\rm phot,1.7}$ & $v_{\rm phot,3.7}$ \\
    (yr) & (au) & (au) & (K) & (K) & (km~s$^{-1}$) & (km~s$^{-1}$) \\
    \hline
     12    & 126 $\pm$ 9  & 138 $\pm$ 6 & 512 $\pm$ 50 & 591 $\pm$ 84 & 63 $\pm$ 6 & 77 $\pm$ 5\\
     14    & 147 $\pm$ 9  & 167 $\pm$ 8 & 497 $\pm$ 42 & 584 $\pm$ 65 & 62 $\pm$ 5 & 80 $\pm$ 7 \\
     44    & 450 $\pm$ 14 & 541 $\pm$ 38 & 378 $\pm$ 22 & 410 $\pm$ 24 & 58 $\pm$ 3 & 72 $\pm$ 3 \\
     \hline
    \end{tabular}
    \caption{The mean stellar radius, temperature and expansion velocity (with their standard deviations) at the photosphere, at three indicative times of the 1.7 and 3.7~\Msun\ simulations. The averages and standard deviations were estimated from a collection of multiple rays launch from the central core.}
    \label{tab:photosphere}
\end{table*} 

This calculation clearly reveals the expansion of the photosphere over time and its morphological changes, as illustrated in Figure~\ref{fig:photo-xsec} and reported in Table~\ref{tab:photosphere}.
To analyse the asymmetry of the flow, we estimate the  radius of the photosphere in the orbital and polar direction. 
To calculate these values, four rays are averaged at each time step to measure the radius in the orbital plane ($\pm$x and $\pm$y) and two rays ($\pm$z) to measure the polar radius, where the radius along each ray is determined in the same way as in Figure~\ref{fig:photo-xsec}.
The average of the four or two measurements on the orbital plane or polar direction are plotted as blue or orange lines in Figure~\ref{fig:photo-size-vs-sep}, along with the average of the photospheric smoothing length, $h_{\mathrm{ph}}$, as an indication of the uncertainty (plotted as a light blue or orange shaded region).
We note that in Figure~\ref{fig:photo-size-vs-sep} (left panel) the size of the photosphere at time zero for the 1.7~\Msun\ model is different between the dusty and non-dusty models. This is due to a slightly different initial model being used in the two simulations, where in the latter the star was relaxed in isolation for slightly longer and expanded to $\sim$2~au. 

Figure~\ref{fig:photo-size-vs-sep} shows that, for the dusty models, the equatorial photosphere rapidly grows to a large size approximately one year before the polar photosphere, effectively indicating that there is a ``hole" at the poles where an observer would see deeper into the object. However, this hole is filled rapidly and the size of the photosphere in the two orthogonal directions becomes very comparable, indicating an approximately spherical object as also seen in Figure~\ref{fig:photo-xsec}. Interestingly, the delay is longer for the 3.7~\msun\ models than for the 1.7~\msun\ one. In the latter a slightly oblate shape is observed in the perpendicular cut until 14.6~yrs, but later in time, at 44~yrs, the shape has become clearly prolate. The effect of these relatively small departures from spherical symmetry on the light will be explored in a future work.

\section{Discussion}
\label{sec:Discussion}

\subsection{Limitations of the current models}
\label{subsec:limitations}

A major shortcoming of the current simulations is the absence of a proper treatment of radiative transfer. Also, contrary to our assumption, the gas and dust temperatures are likely different and these differences will induce heat transfer from one component to the other, thereby introducing cooling/heating terms in the energy equations which are presently not accounted for. 

\citet{Bowen1988} showed that if the equation of state is adiabatic, dust-driven AGB winds have lower mass-loss rates than in simulations with radiative cooling (modelled using an isothermal gas). Including cooling can allow the gas to reach  condensation temperatures in regions closer to the surface of the AGB star and may lead to a larger dust nucleation rate, as dust seeds would form in denser regions. The increase in opacity could in turn enhance the radiative acceleration on the dust, resulting in additional envelope expansion.

Our treatment of the radiative acceleration also assumes that the star keeps a constant luminosity and that photons are not absorbed until they reach the dust grains. Taking into account changes in the luminosity and photon absorption along the ray can impact the flow dynamics, the extent of which still needs to be assessed.  

Another limitation is the lack of grain destruction mechanisms and the account of gas-to-dust velocity drift. In this regard, our simulations show that the densities in dust forming regions are sufficiently high for gas and dust to be dynamically coupled. 
On the other hand, our simulations showed that overdensities from spiral shocks provide sites for dust nucleation, but it remains to be confirmed that this effect persists when including dust destruction mechanisms. 

We currently cannot calculate the nucleation and growth of oxygen rich dust.  As such, we cannot consider dust formation in RGB stars, in oxygen-rich AGB stars, or in massive red supergiants ($\gtrsim 10$~\Msun). We are currently developing a chemical network for the production of oxygen rich dust to address these cases.

\subsection{Comparison with other models of dust production in CE interactions}

\cite{Glanz2018} determined the expected dust condensation radius and dust driving properties from the CE interaction of an RGB star with an main sequence companion. They used analytical models for dust-driving winds in single stars, adopting the temperatures and densities from a non-dusty 3D hydrodynamic simulation.
We use their formula (equation 5), in which they assume that the optical depth of the region below the condensation radius is large, to determine the dust condensation radius, $R_\mathrm{cond}$. We take a condensation temperature of $\sim$1400~K (as indicated by Figure~\ref{fig:dustySPH-particles-T-vs-pressure}), and find $R_\mathrm{cond}\sim1000$~\Rsun ($\sim5$~au) for the AGB star of 1.7~\Msun{} (using a stellar radius of $R=260$~\Rsun) and an effective temperature of 3130~K.

Our simulations indicate that dust forms at distances of about 10--20~au (which is also the distance at which dust starts to form in the post-processed models of \citet{Iaconi2019a}, who, incidentally, use the same star as \citealt{Glanz2018}). Hence, the analytical results of \citet{Glanz2018} indicate a smaller dust nucleation radius for an optically thick envelope.
Furthermore \citet{Glanz2018} did not consider recombination energy in their models.
On the contrary, in our simulations recombination energy is included, and none of it is released. Any number of differences can arise by these two extreme situations, in terms of the expansion and cooling of the envelope, which can alter the time and place of dust formation.

\citet{MacLeod2022} collected observations of luminous red novae analysed and measured by \citet{Matsumoto2022}, and estimated the radius of the dust photosphere at the time when the dust layer becomes opaque, as a function of the mass of the progenitor. They used the time it takes for the {\it optical} lightcurve to dim to 90~per cent of the maximum light, interpreted as the timescale over which 90~per~cent of the transient's optical energy is radiated away and known as $t_{90}$ in \citet{Matsumoto2022}. To calculate the radius, \citet{MacLeod2022} multiplied $t_{90}$ by the velocity of the material obtained from Doppler-broadened emission lines. For their 2 and 4~\Msun\ cases, they find a photospheric radius $r_{\rm 90}\sim 4$ and $\sim 10$~au, respectively. 

Our values, taken $\sim$3~yrs after the start of the simulation when the dust photosphere first appears, are $\sim$32 and $\sim$20~au, for the 1.7 and 3.7~\msun\ simulations, respectively. At this time the dust photosphere is not fully formed for the 3.7~\msun\ model (Figure~\ref{fig:photo-xsec} - top right panels), but the engulfing of the core particles by the dust photosphere is complete shortly thereafter.  
Our estimates of $r_\mathrm{90}$ are somewhat larger for both models, and do not show a clear dependence on the primary's mass, contrary to the tendency towards larger photospheres in more massive stars, as measured by \citet{MacLeod2022}. This may be due to the low mass ratio $M_2/M_1 = 0.16$ (Column 3 of Table~\ref{tab:simulations}) for the 3.7~\Msun\ model, which may result in a relatively smaller and slower dust production.

The amount of dust formed in our simulations ($0.8-2.2\times$10$^{-2}$~\Msun) is similar to the most extreme case calculated by \cite{Lu2013}. They calculated (oxygen) dust formation for common envelope interactions between stars with mass between 1 and 7~\Msun\ both at the base and at the top of the RGB, with a C/O number ratio equal to 0.4 (i.e., with mainly olivine- and pyroxene-type silicate dust grains) and with a 1~\Msun\ stellar companion. They used a dust formation code and a 1D dynamical model of the expanding and cooling ejected CE. Only their RGB tip model with the steepest dependence between temperature and radius achieves a dust production similar to the dust mass produced by our models (which is close to the maximum possible).

\subsection{Dust yields in CE interactions and in single AGB stars}

\citet{Ventura2014} estimated that the amount of carbon dust formed in single AGB stars of 2~\Msun\ and Z=0.008 ranges between $10^{-4}$ and $10^{-3}$~\Msun, depending on the mass-loss prescription. The dust mass can reach $10^{-2}$~\Msun\ at Z=0.004. For a 2.5~\Msun{} Z=0.008  star, they find a dust production ranging between $10^{-3}$ and $10^{-2}$~\Msun. Their values are comparable but somewhat lower than ours ($8.4\times 10^{-3}$ and $2.2\times 10^{-2}$~\Msun, for the 1.7 and 3.7~\Msun\ models, respectively), but we warn that their metallicities are 2.5 and 5 times lower than ours, a parameter that had a large impact on dust production.

The CE interaction interrupts the AGB evolution before its natural end. The C/O ratio of AGB stars increases with time along the AGB. Depending on the mass of the star this can become larger than unity. If the mass is high enough eventually the C/O ratio decreases again due to hot-bottom burning of carbon to nitrogen. Hence interrupting the AGB evolution can result in a different chemistry out of which to form dust. Therefore the dust composition and opacity in a CE interaction may differ from those of the same star, had it been evolving as single.
 
\subsection{Possible observational counterparts}

\citet{Gruendl2008} detected a number of carbon stars in the Large Magellanic Cloud galaxy with very large infrared excesses, SiC in absorption, and an implied mass loss rate of 10$^{-4}$~\Msun~yr$^{-1}$, about ten times larger than expected for single stars. These stars also point to low main sequence masses of 1.5-2.5~\Msun. Galactic counterparts have similar properties, including similar mass-loss rates despite the higher metallicity of the Galaxy. 
\citet{DellAgli2021} analysed a similar dataset and proposed that extreme carbon stars have a CE origin.
CE ejections take place over a short time: the AGB envelopes of the 1.7 and 3.7~\Msun\ models have masses of $\sim$1.2~\Msun\ and $\sim$3.0~\Msun, respectively and are ejected in approximately 20~yrs (Figure~\ref{fig:separation_massunbound}), giving mass loss rates of $\sim$10$^{-2}$--10$^{-1}$~\Msun~yr$^{-1}$. Our simulations may not correctly mimic the immediate post-in-spiral timescale, and are unable to follow the system for the 2-3 centuries over which thermal relaxation may take place. During that time the mass-loss rate would decrease compared to the initial in-spiral values, but it is clear from the simulations that mass-loss rates would be substantially higher than for the inferred single star values. Taken together with the formation of $\sim$10$^{-2}$~\Msun\ of dust on very short timescale it makes the possibility that extreme carbon stars are caught in the aftermath of the CE in-spiral. \citet{Groenewegen2018} list stars with L=5000-10\,000~\Lsun\ and effective temperatures of $\sim$300~K, which would imply photospheric radii between $\sim$100 and 200~au, not dissimilar to our inferred photospheric radii.  It is not a stretch to consider these stars direct outcomes of CE interactions between low mass AGB stars and a companion.

In a similar way, water fountains have been proposed by \citet{Khouri2021} to be post-CE systems. Water fountain stars are oxygen-rich AGB stars, with H$_2$O maser emission related to fast polar outflows. They have optically thick, O-rich dusty envelopes, distributed in a torus shape, and are deduced to have extremely high mass-loss rates of $\sim10^{-3}$~\Msun~yr$^{-1}$, two orders of magnitude higher than the single star prediction. The water fountain phenomenon is estimated to last a few hundred years at most. These objects are compatible with the idea of a fast ejection of gas and dust, resulting in a transitional, but very optically thick dust shell. The seeming conflict between the equatorial dust tori in water fountain sources and the more spherical dust distribution in our simulations needs further study. 
The fast collimated outflows ensuing from the core are possibly magnetically collimated jets due to accretion of modest amounts of fall-back material onto the companion or core of the primary (assuming the binary has survived the common envelope).
This type of fall-back and jet formation has been studied both observationally and theoretically by \citet{Tocknell2014} and \citet{Nordhaus2006} and the mechanisms are completely aligned with the idea that in water fountains we are observing the onset of the jets.

The expected high and fast dust production in CE evolution with AGB stars should leave some distinctive observational signatures in planetary nebulae from CE interactions, compared to nebulae deriving from single star and wider binary interactions. This is not readily observed, possibly due to the complex series of physical phenomena that follow the AGB envelope ejection and which can complicate the observations.
Planetary nebulae do not just form at the moment of the ejection of the AGB envelope (over tens of thousands of years as in single AGB stars or just decades as in common envelope interactions), but by the sweeping and ionising action of the stellar wind that follows the ejection. This may complicate the interpretation of the ejection history by studying the morphology, even in extreme cases, such as CE interactions.

Finally, there are many other possible CE ejections that are only now being discovered in the transient sky. An example are the ``SPRITEs" \citep[eSPecially Red Intermediate-luminosity Tansient Events;][]{Kasliwal2017b}, that may be CE ejections involving more massive, red supergiants. These optically-invisible, mid-IR transients bear the hallmark of copious fast ejection of gas and dust. We leave it to future work to analyse these objects.
\section{Conclusions}
\label{sec:Conclusions}
We have carried out two common envelope simulations with a 1.7 and a 3.7~\msun\ AGB stars with a 0.6~\msun\ companion, including a self consistent treatment of dust nucleation. We have used a C/O ratio of 2.5 to test not only the dust formation properties but also the wind-driving properties. Our main conclusions are:

\begin{enumerate}
\item In both simulations dust formation starts in a shell at about 2-3~yr from the beginning of the simulation, during the early in-spiral. At 5 years the dust forming shell is thin, has irregular thickness, and is located at $\approx 40$~au. For the 1.7~\Msun\ (3.7~\Msun) simulation, this shells moves outward and by 20 years it reaches a distance of $\approx 160$~au (200~au) and is $\approx 40$~au (20~au) thick.

\item Dust formation starts in small grains to which monomers are added, increasing the dust grains size. As the dust grains move away from the seeds formation region, transported out by the expanding envelope, their size stops increasing. The dust that forms earlier ($\approx$2~years) remains smaller ($\approx$0.06~$\mu$m) than dust formed later (between 9 and 15 years; 0.4-0.6~$\mu$m). 

\item The total dust mass is similar in the lower and higher mass simulations up to $\sim$22 years. At later times, the total dust mass is higher for the more massive envelope, eventually plateauing below the theoretical maximum, with total dust yields of $8.4\times10^{-3}$~\Msun\ (1.7~\Msun) and $2.2\times10^{-2}$~\Msun\ (3.7~\Msun). Therefore, the amount of dust formed depends to a greater extent on the metallicity and mass of the AGB envelope, other factors being of lesser relevance. Dust formation is very efficient  ($100$\%), if we do not take into account any dust destruction process.

\item Dust formation does not lead to substantially more mass unbinding, although $\sim10^{-2}$~\msun\ of dust is produced. 
Dust forms too far in the wind to generate effective driving on a dynamical timescale required to aid with envelope ejection, because at those distances, the radiative flux is greatly diluted and gas is already unbound. It is possible --- indeed likely --- that given more time even small accelerations may build larger velocities. 

\item The amount of dust that forms in the common envelope greatly impacts the optical appearance of the model. A simple calculation of the photospheric size of the models shows that at 12 years the photosphere of the dusty models is 4-7 times larger than for the non-dusty models. By 44 years the dusty models are $450-540$~au, and the corresponding mean photospheric temperatures are $\sim$400~K.

\item The morphology of the envelope in the presence of dust formation is nearly spherical. Initially the shape of the envelope is clearly elongated along the equatorial plane for both models. For the 1.7~\msun\ model it remains somewhat elongated (oblate) to $\sim$15~yrs but by 44~yrs it becomes somewhat prolate. For the 3.7~\msun\ model, which has a relatively lighter companion, the shape becomes nearly spherical by $\sim$15 years and remains so till the end of the simulation. 
\item These simulations add further evidence to the suggestion that extreme carbon stars, water fountain stars, as well as some infrared transients might be objects caught in the immediate aftermath of a CE interaction on the AGB. 
\end{enumerate}

\section*{Acknowledgements}

OD, LS, MK and LB acknowledge support through the Australian Research Council (ARC) Discovery Program grant DP210101094. LS is senior research associates from F.R.S.- FNRS (Belgium).  TD is supported in part by the ARC through a Discovery Early Career Researcher Award (DE230100183). DJP is grateful for ARC Disovery Project funding via DP220103767.
CM acknowledge support through the Australian Government Research Training Program Scholarship and acknowledge the use of the python package {\sc Sarracen} \citep{Harris2023} used to post-process {\sc phantom} data dumps.
This work was supported in part by Oracle Cloud credits and related resources provided by Oracle for Research. This work was performed in part on the OzSTAR national facility at Swinburne University of Technology. The OzSTAR program receives funding in part from the Astronomy National Collaborative Research Infrastructure Strategy (NCRIS) allocation provided by the Australian Government, and from the Victorian Higher Education State Investment Fund (VHESIF) provided by the Victorian Government. Some of the simulations were undertaken with the assistance of resources and services from the National Computational Infrastructure (NCI), which is supported by the Australian Government. This research was supported in part by the Australian Research Council Centre of Excellence for All Sky Astrophysics in 3 Dimensions (ASTRO 3D), through project number CE170100013. 

\section{Data Availability}

The data underlying this article will be shared on reasonable request to the corresponding author. All movies for this article can be found in the following URL: \url{https://tinyurl.com/y455avdj}



\bibliographystyle{mnras}
\bibliography{bibliography}

\appendix 
\section{Numerical consideration: the effects of resolution on dust formation and conservation properties of the simulations}
\label{AppendixB}

The effect of resolution on parameters derived from {\sc phantom} simulations of the common envelope interaction has been thoroughly investigated elsewhere \citep[e.g.,][]{Iaconi2018,Reichardt2019,Reichardt2020,Lau2022,GonzalezBolivar2022}. The only quantities that have never been assessed are those connected to dust mass and dust growth. 

\begin{figure}
    \centering
    \includegraphics[width=1.0\linewidth]{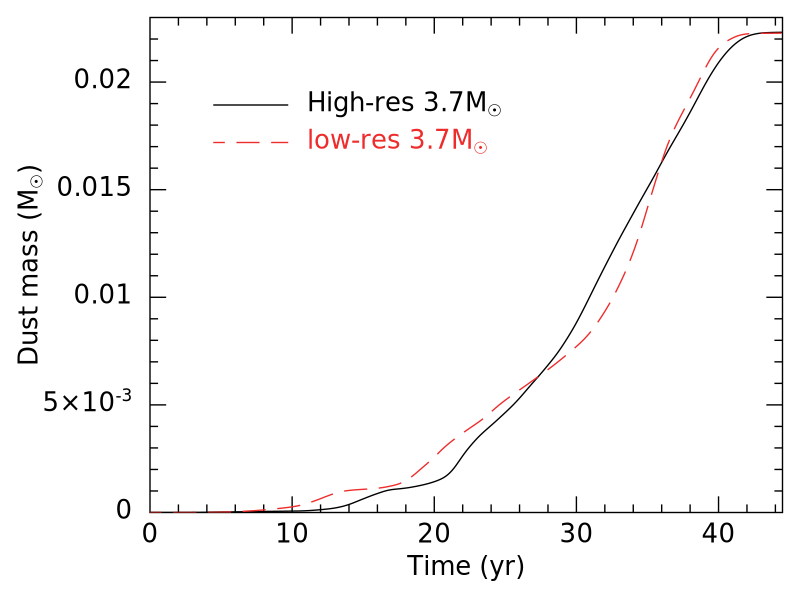}
    \caption{Dust mass as a function of time for the high-resolution 3.7~\Msun\ production simulation presented in this paper ($1.37\times10^6$ SPH particles, black solid line) and a low resolution ($2\times10^5$ SPH particles, red dashed line) identical CE simulation.}
    \label{fig:resolution-test}
\end{figure}

Figure~\ref{fig:resolution-test} displays the evolution of total dust mass with time for two CE simulations with a 3.7~\Msun AGB star, the simulation presented in this paper ($1.37\times10^6$ SPH particles, black solid line) and one with identical stellar parameters but with only $2\times10^5$ SPH particles (red dashed line in Figure~\ref{fig:resolution-test}). From Figure~\ref{fig:resolution-test} it is evident that the amount of dust mass is very weakly dependent on resolution, and at the end of the simulations, the total amount of dust is the same. We therefore suggest that the simulated process of dust formation is nearly-independent of numerical resolution. While it would be appropriate to carry out a more canonical convergence test (at least three simulations differing in number of particles by a factor of at least 8, this is precluded by very long computation times). 

At lower resolution the stellar structure used to carry out the simulation is not as stable as at high resolution. We have checked the stability and behaviour of the single stellar structure at low resolution by evolving it for 5 years in isolation, but doubling the softening length of the core (to 16~\Rsun). While we can confirm that the structure is not as stable, by 5 years only 3~per cent of the particles have expanded past the initial radius and the density profile is not overly distorted. Finally, no dust has formed in the isolated structure.

\begin{figure*}
    \centering
    \includegraphics[width=0.8\linewidth]{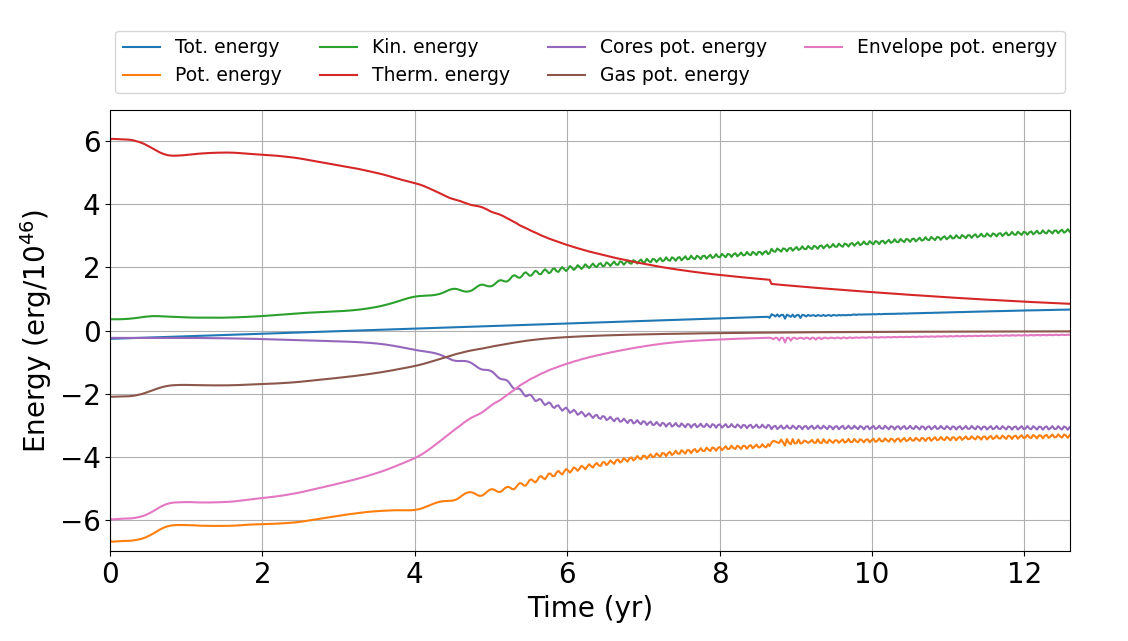}\\
    \includegraphics[width=0.75\linewidth]{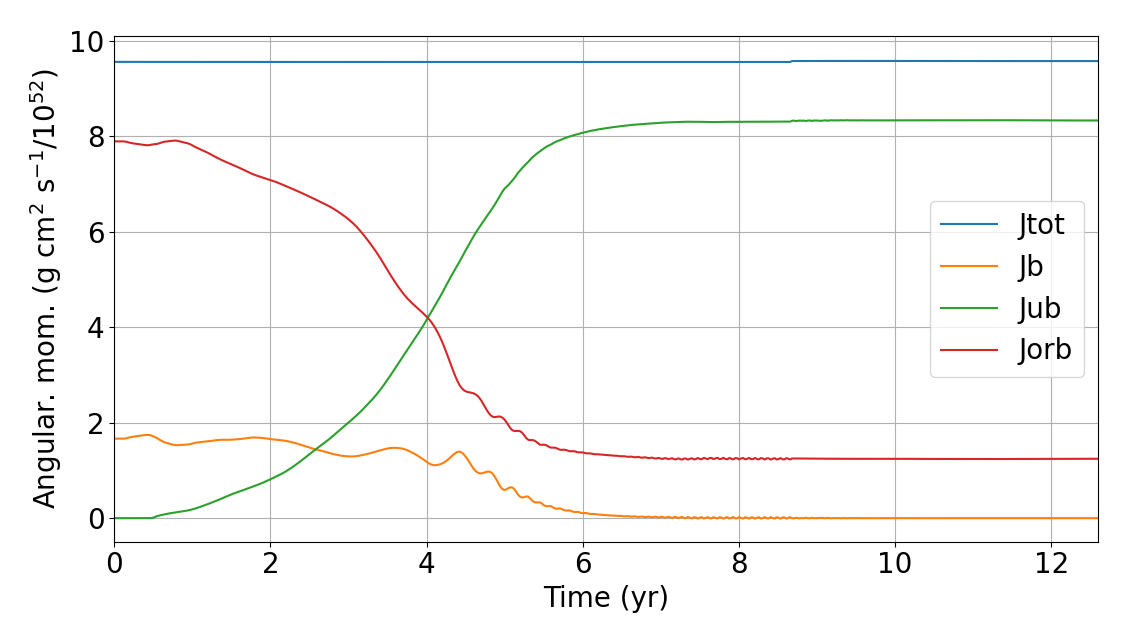}\\
    \caption{Energy (top panel) and angular momentum (bottom panel) evolution for the 1.7~\Msun\ simulation. Top panel: total energy (blue line), total potential energy (yellow line), kinetic energy (green line), thermal energy (red line), potential energy of the point mass particles (violet line), potential energy of the gas without point mass particles (brown line) and potential energy of the gas and point mass particles (pink line). Bottom panel: total angular momentum (blue line), orbital angular momentum (red line), angular momentum of the bound gas (yellow line) and of the unbound gas (green line).}
    \label{fig:energies-momentum-1.7}
\end{figure*}

\begin{figure*}
    \centering
    \includegraphics[width=0.8\linewidth]{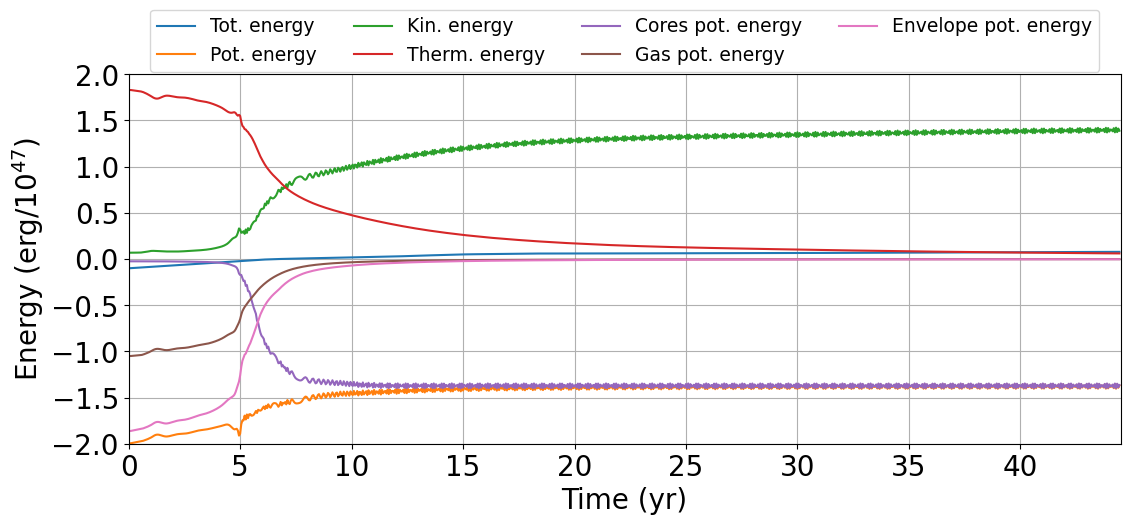}
    \includegraphics[width=0.83\linewidth]{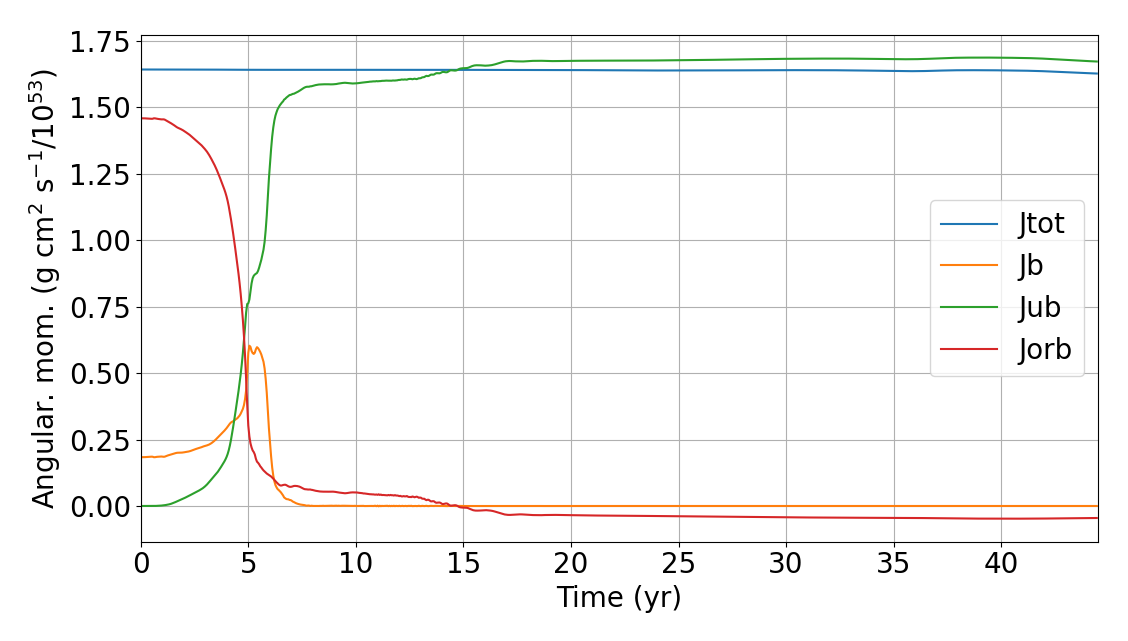}\\
    \caption{Energy (top panel) and angular momentum (bottom panel) evolution for the 3.7~\Msun\ simulation. Top panel: total energy (blue line), total potential energy (yellow line), kinetic energy (green line), thermal energy (red line), potential energy of the point mass particles (violet line), potential energy of the gas without point mass particles (brown line) and potential energy of the gas and point mass particles (pink line). Bottom panel: total angular momentum (blue line), orbital angular momentum (red line), angular momentum of the bound gas (yellow line) and of the unbound gas (green line). }
    \label{fig:energies-momentum-3.7}
\end{figure*}

In Figures~\ref{fig:energies-momentum-1.7} and \ref{fig:energies-momentum-3.7} we evaluate the conservation and redistribution of energy and angular momentum in the dusty CE simulations. In the top panel the total energy (blue line) is conserved to within $1\%$. As the envelope expands the thermal energy decreases (red line), as does the potential energy of the cores that in-spiral towards one another (purple line). Every other energy (kinetic energy of the cores and gas (green line) potential energy of the envelope gas and cores (pink line) and potential energy of the gas (brown lines). (Note the orange line is the total potential energy, core-core, core-gas and gas-gas).

Energy conservation is achieved despite the fact that we inject energy in the form of radiative accelerations. The fact that the effect of the last term in Equation~\ref{eq:motion} is negligible is a result of the comparatively small accelerations that result from it, as explained in Section~\ref{ssec:orbital-evolution}.

In the bottom panels of Figures~\ref{fig:energies-momentum-1.7} and \ref{fig:energies-momentum-3.7} we plot  the total angular momentum (blue line) that is conserved with an error of less than $1\%$. As the companion plunges into the primary's envelope, the orbital angular momentum (red line) decreases and is transferred to the bound (yellow line) and unbound gas (green line). 
After 7 years, the inner gas region is almost in corotation, the gravitational drag force is considerably reduced, the separation stabilises and the orbital angular momentum becomes almost constant. From that moment there is no more angular momentum transfer from the point mass particles to the gas. The angular momentum of the bound material decreases as the amount of bound material is  reduced due to recombination  energy conversion into thermal energy and ultimately into work. 

\bsp	
\label{lastpage}
\end{document}